\documentclass[preprint]{elsarticle}
\usepackage[english]{babel}
\usepackage{array}
\usepackage{placeins}
\usepackage{ragged2e}
\usepackage{tabularx}
\usepackage[table]{xcolor}
\newcolumntype{L}[1]{>{\hsize=#1\hsize\raggedright\arraybackslash}X}%
\newcolumntype{R}[1]{>{\hsize=#1\hsize\raggedleft\arraybackslash}X}%
\newcolumntype{C}[1]{>{\hsize=#1\hsize\centering\arraybackslash}X}%
\newcolumntype{Y}{>{\raggedright\arraybackslash}X}
\usepackage{lineno,hyperref}
\usepackage{amssymb}
\usepackage{amsmath}
\usepackage{ulem}
\usepackage{interval}
\usepackage{pgfkeys}
\intervalconfig{soft open fences, separator symbol=; ,}
\usepackage{cases}
\usepackage{threeparttable}
\usepackage{enumitem}
\usepackage{booktabs,multirow}
\usepackage{makecell}
\newcommand{\lra}{\Leftrightarrow}
\newcommand{\uda}{\Updownarrow}

\usepackage{upgreek}
\usepackage{mathtools, cuted}
\usepackage{lipsum}
\usepackage{textcomp, gensymb}
\usepackage{gensymb}
\usepackage{color}
\usepackage{graphicx}
\usepackage{wrapfig}
\usepackage{subcaption}
\usepackage{float}
\usepackage{natbib}

\newcommand{\vpb}{\vphantom{\bigg|}}

\usepackage{geometry}
\biboptions{sort&compress}

\begin{document}

\begin{frontmatter}

\title{Analytical Prediction and Numerical Verification of Stress Concentration Profile Around an In-situ Tow Break in Resin-impregnated Filament-wound Composites}

\author[cumae]{Jiakun Liu}\corref{cor1}\fnref{upenn}
\ead{jl2938@cornell.edu}
\author[cumae]{Stuart Leigh Phoenix}
\ead{slp6@cornell.edu}

\cortext[cor1]{Corresponding author}
\fntext[upenn]{Current affiliation: Department of Mechanical Engineering and Applied Mechanics, University of Pennsylvania, 220 S 33rd St, Philadelphia, PA 19104, USA}

\address[cumae]{Sibley School of Mechanical and Aerospace Engineering, Cornell University, Ithaca, NY 14853, USA}

\begin{abstract}
A new empirical analytical approach is developed for predicting the stress concentration profile around an in-situ tow break in filament-wound composites. A shear-lag analysis is firstly performed to solve for the perturbational axial displacement of the broken tow and resultant debonding lengths. Solution of stress field caused by tangential load on the surface of a elastic half space is then utilized in combination with superposition concepts to obtain the overload magnitudes in the neighboring tows. Subsequently, high-fidelity finite element analysis on a representative uni-directional laminate model under different stress states is performed, and excellent overall agreement is observed between analytical and numerical predictions. The proposed method takes into account essential aspects such as transversely isotropic material properties, in-situ stress states and their effect on the interfacial frictional forces in the debonded interfaces, and thus provides a convenient way to evaluate the stress concentration factors in damaged filament-wound composites. In addition, this approach can be applied to yield auxiliary failure evaluation criteria for statistical strength prediction or finite element modeling of filament-wound composites or similar structures.
\end{abstract}

\begin{keyword}
Filament-wound Composites
\sep Resin-impregnated Tows
\sep Stress Concentration Factor
\sep Finite Element Modeling
\end{keyword}

\end{frontmatter}

\section{Introduction}\label{sec:intro}

\subsection{Challenges in damage analysis of FWCs}
\begin{figure}[!htb]
\centering
\includegraphics[width= 0.75\textwidth]{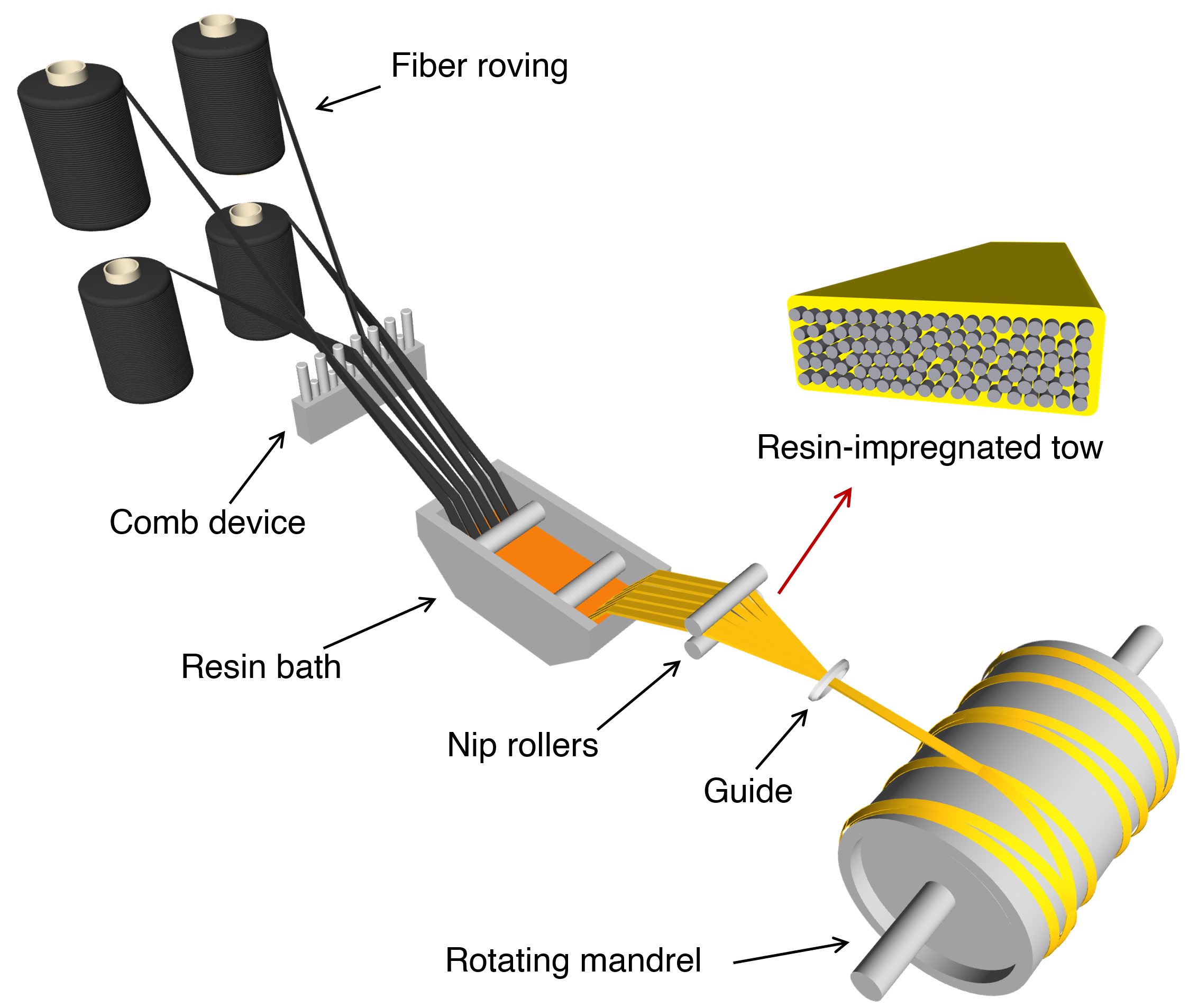}
\caption{Basic procedures for resin-impregnation of carbon fiber tows and filament winding process (this plot shows wet-winding).}
\label{fig:Filament-winding}
\end{figure}

The filament winding technique is efficient, cost-effective, and has been applied in fields including civil and environmental, aerospace, and automotive industries. It is primary used to manufacture hollow or convex-shaped components such as wind turbines, pipes, gas tanks, and composite overwrapped pressure vessels (COPVs). Fig.\ref{fig:Filament-winding} illustrates basic procedures of filament winding, during which combed fiber strands are impregnated in hot resin bath, and then the resin-impregnated tows are wound onto a rotating mandrel to fabricate the filament-wound composites (FWCs). Filament tows are usually impregnated right before being wound onto the mandrel (wet-winding), but can also be priorly-impregnated (dry-winding), or post-impregnated. During winding, a moving guide (pay-out eye) wraps filament bands onto a rotating mandrel following a layer-over-layer routine. The motion and path of the guide are controlled by computer-aided winding machines to help realize desired layer angle at specific positions and targeted thickness of each winding layer. After completion of winding, the entire structure is placed in an oven or under radiant heaters for curing. Thereafter, the mold core may be removed or left as an integral component of the final product. 

Since FWC structures are widely used in engineering applications, and their failure could lead to devastating losses (e.g., burst of COPVs), it is of critical importance to investigate the damage and load redistribution mechanisms in FWCs. However, these remain as challenging tasks due to the complicated material and structural characteristics \cite{Liu.Zheng2010Review, Zhang.etal2019Review, Li.etal2023Review}.

The first challenge in studying failure mechanisms of composites is associated with the length scales. Composite materials typically exhibit a strong dependence on varying length scales, commonly categorized into: micro, meso, and component scales. Micro scale is generally in the magnitude of microns due to the typical fiber diameters (e.g., $3$ to $10\mu\text{m}$). Meso scale usually ranges from several to hundreds of millimeters depending on the object of interest, and a typical meso-scopic representative volume element (RVE) is used to show the geometrical features. For instance, a periodic RVE that demonstrate the morphology of woven textile composites. Component scale is on the magnitude of a structure or part (e.g., composite wing blades) for structural analysis. This multi-scale analysis concept is important because certain features may only be analyzed or modeled at a specific length scale. Also, due to the complicated geometries, material heterogeneity, and enormous number of fibers in composite structures, direct linking and information sharing between scales is impractical. As a result, computational homogenization and multi-scale modeling methods are widely applied for analysis of composites.

In addition to length scale, another challenge is the morphological complexity of FWCs. Firstly, the local pattern of filament can be highly intricate. For example, overwrap of a cylindrical COPV usually has both both helical and hoop layers, each with specific winding angles and layer thicknesses, and henceforth, the stacking angles of tows vary among different locations and also between distinct layers. Secondly, the material integrity and distribution are not homogeneous. For example, raw roving strands (used as source material of filaments) can be regarded as carbon fiber tows embedded in dry matrix, and as these tows pass through hot resin flows, additional wet resin coatings are adhered to the strand surfaces (shown schematically in Fig.\ref{fig:Tow-CrossSection}), resulting in resin-rich areas among tows and between different layers. 

In a mechanical perspective, unevenly distributed fiber volumes lead to 'weak interfaces' affecting local stress redistribution and damage paths, which, in some cases, could dominate the strength and failure mechanisms of composites. Such kinds of heterogeneity and resin-rich areas can be evidently observed in both intact and damaged filament-wound and/or prepreg-fabricated composites \cite{Grunenfelder.etal2013, Cherniaev.Telichev2016, Schechter.etal2020}. Meanwhile, discrete failure patterns in form of split strands/tows can be observed in widely available photos of burst/destructed FWCs \cite{Zheng.Liu2008, Wang.etal2016, Alam.etal2021, McLaughlanP.B..Grimes-LedesmaL.R.2011, Almeida.etal2016, Zu.etal2019, Liu.Phoenix2019}, where interfacial cracks among broken tows and 'brush-like' damage patterns (split and protruding tows) are commonly observed failure modes. These phenomena are expected considering the fact that initial raw materials used for filament wounding process are in the formation of such 'tows' or 'strands'.

\subsection{Theoretical strength prediction of fibrous composites}

To provide theoretical foundation for strength prediction of FWCs (e.g., COPVs), or more generally, fiber-reinforced polymeric composites, an extensive amount of studies on stress redistribution mechanisms between (micro-scopic) fibers and matrices have been presented in the past decades, such as Refs. \cite{Hedgepeth1961, Hedgepeth.VanDyke1967, Beyerlein.Phoenix1996, Beyerlein.etal1996, Ibnabdeljalil.Curtin1997, Babuska.etal1999, Landis.McMeeking1999, Landis.McMeeking1999a, Mahesh.etal2002, Mahesh.etal2004, Swolfs.etal2015a, Engelbrecht-Wiggans.Phoenix2019}, among many others.  These work typically focus on a fiber-matrix system with assumed load-sharing mechanism among one or multiple fiber breaks in specific geometrical patterns. Through which, the stress concentration factor (SCF) regarding a cluster of fiber breaks (at micro scale) can be quantified, and thereafter, the strength of large volume composites can be predicted by finding a 'critical cluster size' (number of broken fibers) above which the SCF would be large enough to trigger catastrophic failure. This concept is somehow comparable to the 'critical crack length' in classical liner elastic fracture mechanics (LEFM) for strength prediction of brittle materials, above which the resultant stress intensity factor would become significant enough to trigger unstable crack propagation.

When applying the above micro-mechanical fiber breakage analysis results to predict strength of component-scale composites with fiber volume $V$ (here referring to the number of fibers), the entire composite is typically conceptualized as an 'aggregate', or a 'linkage', of $n$ inter-connected micro-scale fiber-matrix bundles, where each bundle contains $m$ fibers so that $V=mn$. 
Then the statistical strength prediction is performed by finding the critical number of fiber breaks (and distributions) and the corresponding probabilities of failure of such a linked fiber bundles \cite{Phoenix.Beyerlein2000, Thesken.etal2009, Pradhan.etal2010}. 
To apply micro-mechanical results, bundle size has to be limited (e.g., $m<20$) so that load sharing mechanisms correlated to all possible fiber breakage amount and patterns in a bundle, and henceforth SCFs in unbroken fibers in the bundle, can be analytically quantified. And all $n$ bundles are subsequently assumed to share similar load-sharing mechanisms. However, in reality, load redistribution among broken and intact fibers may vary among different locations, and may also depend on multiple factors such as winding patterns, in-situ stress states. For instance, in a cross-ply ($\pm \theta^{\degree}$) laminate, assume a cluster of fiber breaks occurred around the top of an inner $+\theta^{\degree}$ ply below a $-\theta^{\degree}$ ply, then the load redistribution mechanisms between fibers in the same ply would be different from the mechanism between fibers across the cross plies.

\subsection{Numerical approaches for failure analysis of FWCs}
Other than analytical methods and experimental studies, finite element analysis (FEA) is nowadays widely applied for damage analysis and strength prediction of FWC structures \cite{Zheng.Liu2008, Xu.etal2009, Wang.etal2016, Hua.etal2017, Canal.etal2019, Pramod.etal2019, Zhang.etal2020, Rafiee.Abbasi2020, Alam.etal2021}, commonly directly at component scale (e.g., a numerical model of a COPV). In which, the composite part is typically treated as a 'smeared' continuum laminate consisted of many winding plies, and the effective properties of such a laminate is obtained by classical laminate theory . Then conventional continuum damage mechanics (CDM) based failure criteria (e.g., max strain, Tsai-Hill, Hashin) or proposed failure evaluation methods with more elaborate algorithms are incorporated with these elements as an 'indicator' of their damage levels. Henceforth the burst strength prediction of FWCs depends on the evaluation of selected failure criteria. However, as mentioned earlier, composites typically exhibit strong scale dependency and characteristics, whereas CDM-based criteria are inherently homogenized evaluations on continuum solid elements with 'smeared' laminate properties. Therefore, despite component-scale structure and loading features (e.g. COPV shape,  effective layer-level stress states) are inherently retained in these kinds of studies, advances in understanding of micro- and meso-scale damage mechanisms would be beneficial in terms of refining in CDM-based criteria, or serving as a failure evaluation method for multi-scale damage analysis.

\subsection{Necessity of damage analysis of a broken tow at meso scale}
In general, the above discussion indicates that a study that covers essential damage mechanisms around a broken tow and also realizes accurate prediction of resultant overload magnitudes is strongly needed. Such a study, on the one side, could provide extra conceptual foundations for statistical strength prediction in analytical approaches that commonly merely rely on micro-mechanical results, and on the other side, could be applied to enhance, or integrate with, existing failure criteria used for component-scale numerical models.

Under such a motivation, in this study we develop a new analytical approach to quantify the overload magnitude around a tow break inside a FWC structure, and then compare analytical results to numerical predictions realized by high-fidelity finite element simulations. The remaining of this article is organized as follows: In Section \ref{sec:shearLagAnalysis}, we start with a shear-lag based analysis that predicts axial deformation of broken tow and interfacial debonding lengths. Then in Section \ref{sec:quantifySCF}, we derive algorithms based on classical contact mechanics (point load analysis) that quantitatively evaluates stress profiles around a broken tow. Subsequently, in Section \ref{sec:feaValidation}, a series of finite element simulations are performed and compared with analytical predictions. Finally, a discussion with concluding remarks are given in Section \ref{sec:conclusion}.

\section{Shear-lag based analysis of an in-situ tow break}\label{sec:shearLagAnalysis}
Many studies involving SCF of fiber breaks applied so-called shear-lag analysis, wherein fibers are considered to be relatively rigid and hence carry all the tensile load, while the epoxy matrix interfaces are the 'load redistributor' that only carry and deliver shear stresses. Earlier work typically assume perfectly-bonded fiber/matrix interfaces even when a fiber breaks \cite{Hedgepeth1961, Hedgepeth.VanDyke1967}. Some later studies included debonding into consideration, wherein the interfacial behavior is commonly described in a 'debond-then-slip' manner \cite{Beyerlein.etal1996, Landis.McMeeking1999, Landis.McMeeking1999a, Xia.etal2001, Mahesh.etal2004}.

In this study, a similar shear-lag concept is used. In specific, it is assumed that the tow breakage, resulted from a critical cluster of fiber breaks inside the tow, will also cause debonding along tow interfaces. Then as the broken tow slides against the surrounding neighbors, interfacial cracks keep propagating until a local equilibrium state has been reached. Such an equilibrium state is when the accumulated shear force, provided by frictional forces in debonded interfaces and elastic shear in deformed interfaces, are sufficient to 'stop' the broken tow and fully recovers its stress after a certain distance. In some conditions, a tow break may become unstable, for example, a broken tow on the outer surface of an COPV may just pump out of the overwrap due to insufficient boundary constraints. In other situations, the overload exerted by tow break(s) may become high enough to trigger continued tow breaks thereby the entire structure will fail catastrophically in a transient amount of time, essentially a typical failure mode (burst) of FWCs.

\subsection{Representative tow cross section}
\begin{figure}[!htb]
\centering
\includegraphics[width=0.5\textwidth]{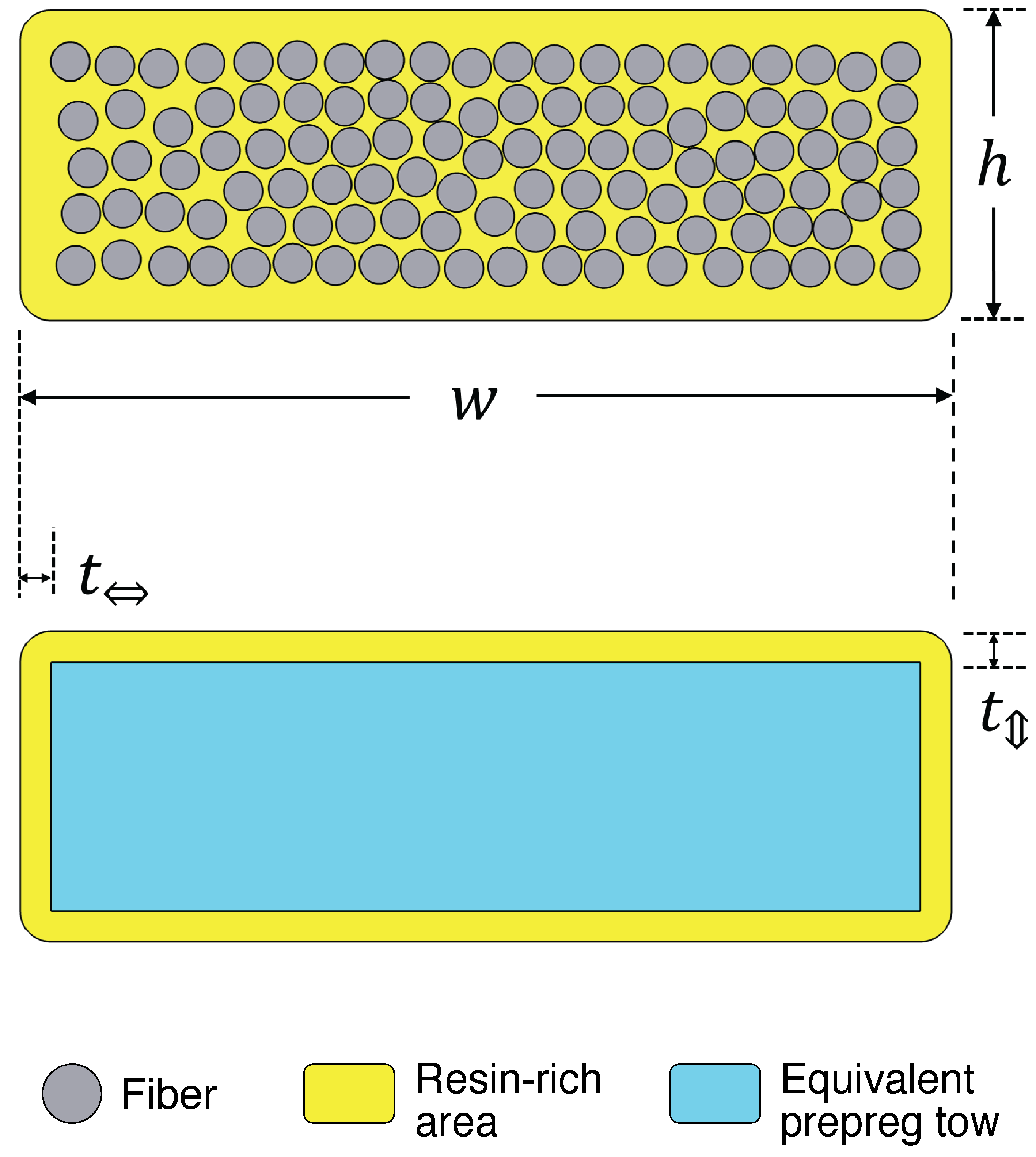}
\caption{Schematic plot of material formation in a resin-impregnated tow and dimensions of its representative cross section.}
\label{fig:Tow-CrossSection}
\end{figure}

A schematic plot of tow cross-section and resin-rich zones is shown in Fig.\ref{fig:Tow-CrossSection}. Unlike in the micro-mechanical analysis where fiber and matrix have comparable dimensions, the resin-rich zone between tows are considered to be much smaller than the cross-sectional dimension of composite tows. The shape of the cross-section is set to be 'almost rectangular' with height $h$ and width $w$, while in practice it could be somewhere between oval and rectangular.

Let $t_{\lra}$ and $t_{\uda}$ be the interface thickness between neighboring tows and adjacent layers, respectively, where the directional arrows used as subscripts are intended to described the interface relations: the left-right arrow $\lra$ indicating the intra-ply property/interaction between tows in the same ply, while the up-down arrow $\uda$ denotes the inter-ply property/interaction across different plies. These subscripts will be used consistently for several other intra- and inter-ply variables in the following contents.

In general, the aspect ratio of tow cross section ranges from $1.0 < w/h < 3.0$, and the cross-sectional area $w \cdot h$ is typically less than $1 \, \text{mm}^2$. The number of fibers within the cross-sectional area of a tow generally ranges from 800 to 2400. For instance, a high modulus carbon fiber strands, Toray\textsuperscript{\textregistered} T1100G, is available in 12k and 24k (density of fibers in tow), based on the fiber diameter ($5\mu\text{m}$) and its typical volume fraction ($60\%$), the cross-sectional area of 12k tow can thus be approximated to be around $0.393 \text{mm}^2$.

\subsection{Coordinate system and notations involving the tow break}
\begin{figure}[!htb]
    \centering
    \includegraphics[width=0.75\textwidth]{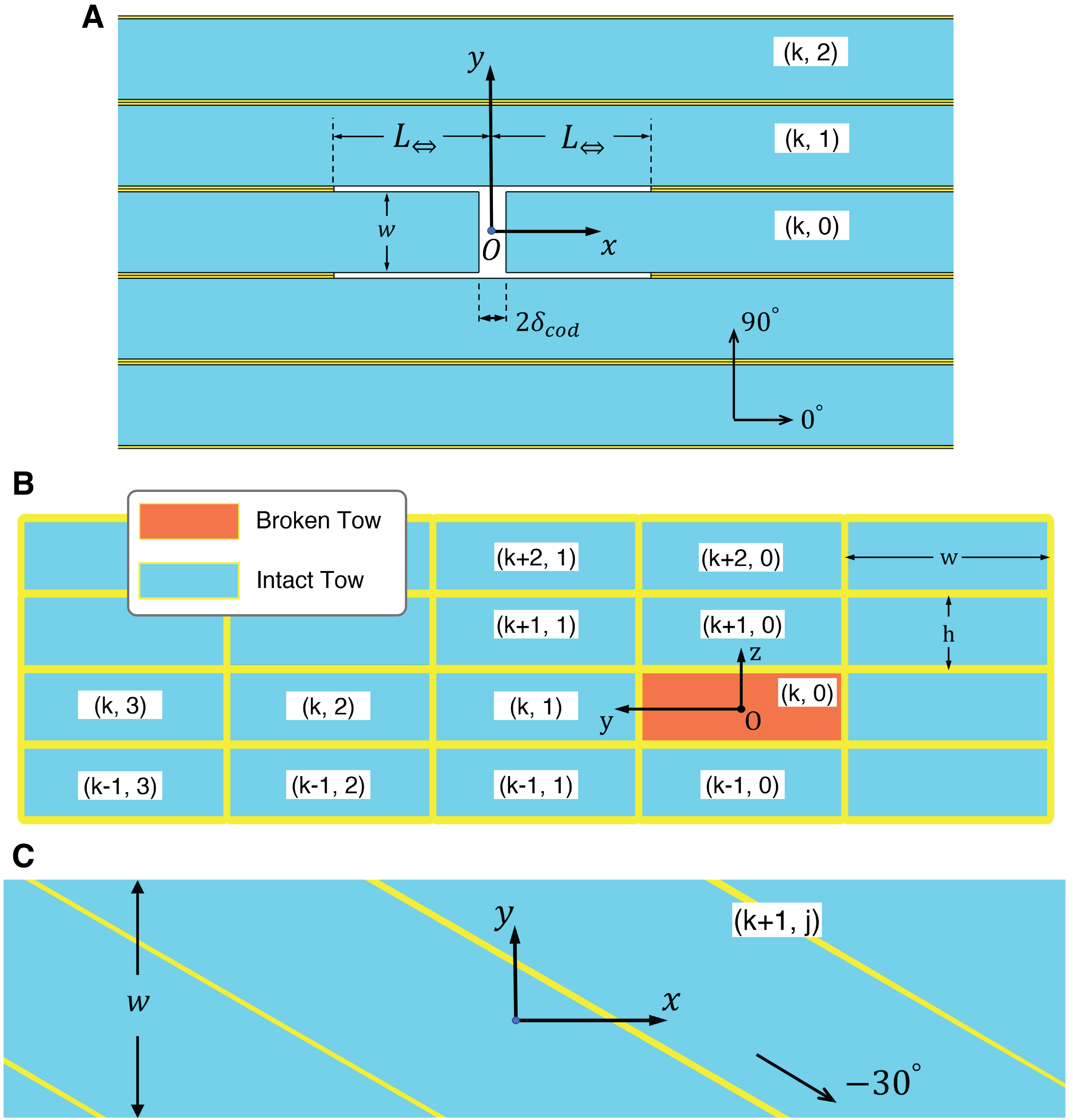}
    \caption{Coordinate system and notations for plies and tows. 
        (A) Intra-ply slice view ($z=0$ plane) around the tow break, showing notations of the neighboring tows in the $k$th ply. 
        (B) Through-thickness slice view in the plane of tow break ($x=0$ plane), showing notations of the neighboring tow regions in different plies.
        (C) If the adjacent plies have different winding angles, the effective tow region may cover a portion of multiple tows. Here as an example, assume the first ply above the broken tow has a winding angle of $-30^{\circ}$, then superscript ($k+1, j$) corresponds to an effective region covering multiple tows.}
    \label{fig:csys-and-notation}
\end{figure}

Consider a small 'piece' cut out of a filament-wound composites, it can then be characterized as a laminate consisted of plies whose stacking layups are equivalent to the winding angles at the cut location. Use x-y-z to denote the global Cartesian coordinate system (CSYS), and use $1{-}2{-}3$ to refer to the local tow material CSYS of each ply, where '$1$' is the longitudinal fiber orientation of tows, '$2$' the transverse direction (vertical to '$1$'), and '$3$' the through-thickness orientation, respectively. Let the winding direction (longitudinal tow orientation '1') of the ply containing the broken tow aligned with x-axis of the global system (Fig.\ref{fig:csys-and-notation}). The winding angle of the $i\text{th}$ ply, $\theta^{i} \in \interval{-90^\circ}{90^\circ}$ denotes the angle between local fiber orientation of the ply to the global x-axis. Let $z_{i}$ ($i = 1,2 ... N$) be the coordinate of bottom surfaces of plies in the $z$ direction, and $z_{N+1}$ be those of the top surfaces of the $N$th ply. The thickness of the $i$th ply is thus $h^{(i)} = z_{i+1} - z_{i}$. These conventions will be used in the remaining contents, where subscripts $x,y,z$ refer to any measure/value of a variable in the global coordinates, subscripts $1,2,3$ refer to measure/value in the local material orientation of a specific ply, and superscript $(i)$ denote those of the $i\text{th}$ ply.

In this study, a primary interest is the stress profile in the neighboring tows right next to the broken tow, since these regions/locations are expected to suffer highest overload. Set the center of this ply as the origin $O$ (Fig.\ref{fig:csys-and-notation}), then following the previously defined superscript $(k)$, let $(k,j)$ be the $j\text{th}$ nearest tow (in the $k$th ply) next to the center tow along the positive $x$ direction, and $(k,0)$ refers to the broken tow itself. Similarly, as shown in Fig.\ref{fig:csys-and-notation}B, denote the tow regions in the above and below plies in the same manner.

Now let the $k$th ply be an inner ply ($2 \leq k \leq N-1$) with winding angle $0^{\circ}$, assume that a tow break occurs in the center of this ply. The dimensions of selected representative laminate are thought to be sufficiently large to cover the stress recovery lengths, so that the stresses or strains at the boundaries are unaffected by the broken tow. Assume that the intra-ply stresses are uniform through the thickness of each ply, whereas the through-thickness stresses may have a slight variation. Denote, respectively, the 'far-field' intra-ply stresses and average through-thickness stress of the ply as
\begin{equation}\label{eq:far_field_stress}
    \sigma^{(k)}_{11,\infty} = \sigma^{(k)}_{xx,\infty}, \quad \sigma^{(k)}_{22,\infty} = \sigma^{(k)}_{yy,\infty}, \quad \bar{\sigma}^{(k)}_{33} = \bar{\sigma}^{(k)}_{zz} = \int_{z_k}^{z_{k+1}} \sigma^{(k)}_{33} dz 
\end{equation}
where the equivalence between local ($1{-}2{-}3$) and global ($x{-}y{-}z$) stresses is due to the assumed $0^{\circ}$ winding angle this ply. It is also noted that the intra-ply shear stress $\sigma^{(k)}_{12}$ of this $0^{\circ}$ ply, under the bi-axial loading, is neglected in the following shear analysis. 

\begin{figure}[!htb]
    \centering
    \includegraphics[width=0.75\textwidth]{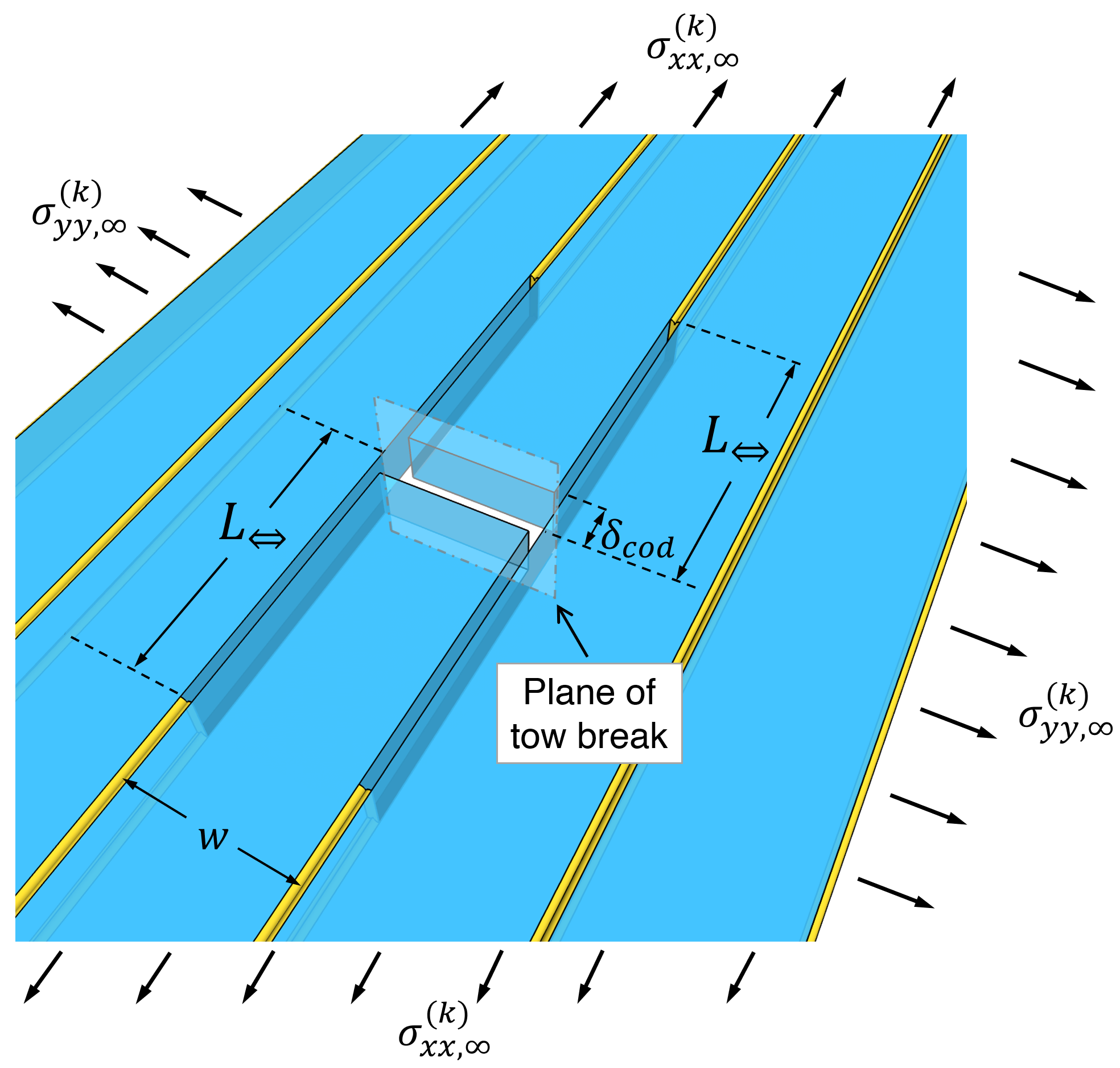}
    \caption{Intra-ply cut-view of an in-situ tow break occurred at center of the $k$th ply with winding angle $\theta^{k}=0^{\circ}$.}
    \label{fig:cutView3D-Intra}
\end{figure}

Fig.\ref{fig:cutView3D-Intra} shows a schematic plot of the broken tow and its intra-ply neighboring tows. The resultant intra-ply debonding lengths, under the in-situ stress states, are $L_{\lra}$ on each side. The distance from the origin to the surfaces of broken tow is $\delta_{cod}$. Since the deformation and debonding lengths of this ply is assumed to be symmetric about the plane of tow break, here $\delta_{cod}$ is half of the real 'gap' between the two fracture surfaces.

\begin{figure}[!htb]
    \centering
    \includegraphics[width=0.75\textwidth]{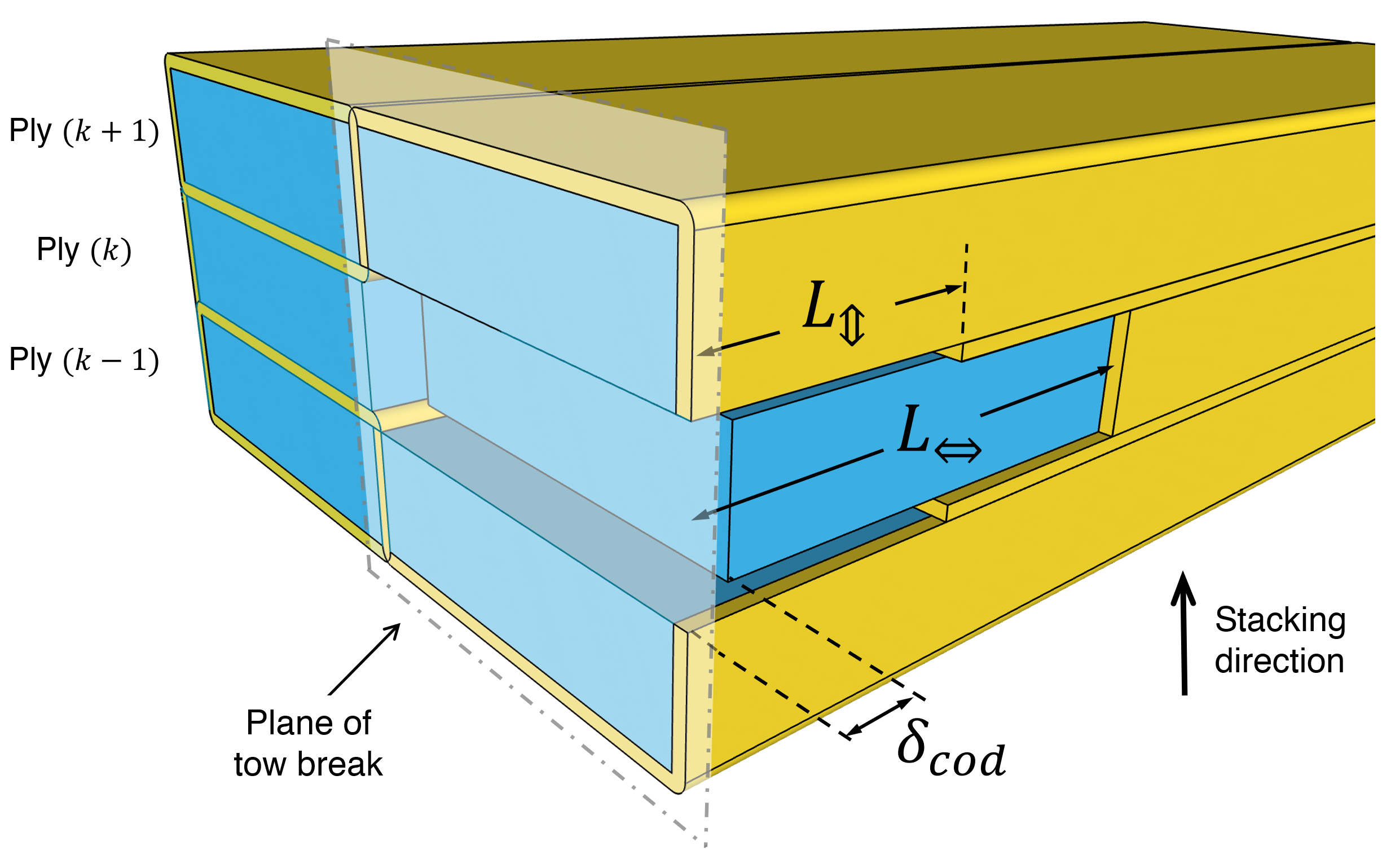}
    \caption{Inter-ply cross-sectional cut-view of an in-situ tow break and corresponding interface debonding.}
    \label{fig:cutVew3D-Inter}
\end{figure}

Fig.\ref{fig:cutVew3D-Inter} shows a schematic inter-ply cross-sectional cut-view of the broken tow and its intra- and inter-ply neighboring tows, where the inter-ply debonding lengths on the top and bottom interfaces of the broken tow are assumed to be $L_{\uda}$. This is based on the assumption that $\sigma_{33}^{(k)}$ does not vary significantly through one ply in the through thickness direction, so that the debonding lengths between the above and below ply interfaces are comparable if these plies have the same winding angles. 

Note that both Fig.\ref{fig:cutVew3D-Inter} and Fig.\ref{fig:cutView3D-Intra} show a representative laminate as a stacking of $0^{\degree}$ tows having the same ply thicknesses. In general, since winding layers can have different angles and thicknesses, the region of interest in the other layers are not necessarily in the form of individual tow/yarns. Instead, such a region may cover multiple tows due to the mismatched winding angles (as illustrated in Fig.\ref{fig:csys-and-notation}C showing a $-30^{\circ}$ layer).

\subsection{Elastic properties and stress states}
Material properties of composite tows are regarded here as linear elastic, brittle, transversely isotropic. Thus the lamina/ply stress-strain relation, in the corresponding local CSYS, is
\begin{equation}\label{eq:stress-strain-local-coord}
    \underset{-}{\varepsilon^{(i)}_{local}} = \underset{=}{S^{}_{tow}} : \underset{-}{\sigma^{(i)}_{local}} 
    \quad \rightarrow \quad 
    \underset{-}{\sigma^{(i)}_{local}} = \underset{=}{S^{}_{tow}}^{-1} : \underset{-}{\varepsilon^{(i)}_{local}} = \underset{=}{C^{}_{tow}} : \underset{-}{\varepsilon^{(i)}_{local}} 
\end{equation}
where the single and double underlines are used to denote first and second order tensors, $\varepsilon^{(i)}_{\,local}$ and $\sigma^{(i)}_{local}$ are the local strain and stress vectors, respectively, ${S}_{tow}$ is the compliance matrix of the tow, and its inverse $C_{tow}$ is the stiffness matrix.

There are five independent variables to be defined: let $E_l$ be the Young's modulus of composite tows in the longitudinal (fiber) direction, $E_t$ be the modulus in transverse directions; $\nu_{lt}$ and $\nu_{tt}$ be the corresponding Poisson's ratios; $G_{lt}$ be the longitudinal-to-transverse shear modulus. Then Eq.\ref{eq:stress-strain-local-coord} takes the form
\begin{equation}\label{eq:stress-strain-tow-matrix-form}
    \begin{bmatrix}
        \varepsilon_{11} \\ \varepsilon_{22} \\ \varepsilon_{33} \\ \gamma_{23} \\ \gamma_{13} \\ \gamma_{12} 
    \end{bmatrix}
    = \begin{bmatrix}
        1/E_l         & -\nu_{tl}/E_t & -\nu_{tl}/E_t & 0        & 0        & 0        \\
        -\nu_{lt}/E_l & 1/E_t         & -\nu_{tt}/E_t & 0        & 0        & 0        \\
        -\nu_{lt}/E_l & -\nu_{tl}/E_t & 1/E_t         & 0        & 0        & 0        \\
        0             & 0             & 0             & 1/G_{tt} & 0        & 0        \\
        0             & 0             & 0             & 0        & 1/G_{lt} & 0        \\
        0             & 0             & 0             & 0        & 0        & 1/G_{lt}
    \end{bmatrix} 
    \begin{bmatrix}
        \sigma_{11} \\ \sigma_{22} \\ \sigma_{33} \\ \sigma_{23} \\ \sigma_{13} \\ \sigma_{12} 
    \end{bmatrix}
\end{equation}
where $\gamma_{23}, \gamma_{13}, \gamma_{12}$ are engineering shear strains, and
\begin{equation}\label{eq:def_Gtt_nutl}
    G_{tt} = \frac{E_t}{2(1+\nu_{tt})}, \qquad \nu_{tl} = \nu_{lt} \frac{E_t}{E_l} 
\end{equation}
are the shear modulus in the transverse plane, and Poisson's ratio in transverse-to-longitudinal directions, respectively. 

Each ply is considered to be in 3D stress states, among which, shear stresses associated with through thickness directions ($\sigma_{23}$ and $\sigma_{13}$) are neglected, namely, the ply-level effective stresses are
\begin{equation}\label{eq:ply_stress_sigma}
    \underset{-}{\sigma^{(i)}_{local}} = \left[ \sigma^{(i)}_{11}, \sigma^{(i)}_{22}, \sigma^{(i)}_{33}, 0, 0, \sigma^{(i)}_{12}\right]^T
\end{equation}
and such an assumption on the ply-level stress state is based on the following considerations: 
1) This is a 'piece' of filament-wound laminate whose boundaries are constrained by the surrounding materials. FWC structures are typically axi-symmetric in geometry and the loading (pressure) applied on the inner surface is also axi-symmetric. And henceforth the shear deformation associated with through thickness directions ($\sigma_{23}, \sigma_{13}$) are considered to be negligible. 
2) Even for a high performance COPV, pressure load through the thickness direction is at least one order of magnitude lower than the intra-ply stresses. Typical burst pressure of commercial COPVs are less than $20,000 \,$psi ($\approx$ 138MPa), which is significantly smaller than the fiber tensile stresses. Therefore the magnitudes of $\sigma_{23}, \sigma_{13}$ are expected to be small as compared to fiber tensile stresses.
3) This study involves in-situ analysis of damages in pressure-bearing structures, so axial stress in the thickness direction, $\sigma_{33}$, cannot be neglected. 

\subsection{Interfacial shear stress and force equilibrium}\label{subsec:CZM}
For composite prepreg tows, $E_l$ is typically one order of magnitude higher than $E_t$ and the modulus of resin. Therefore, similar to the aforementioned micro-mechanical shear-lag analysis, assume that the axial strains and stresses are uniform through the cross section of the broken tow, so that they only depends on $x$ (distance away from the plane of tow break). Then based on the previously established notations, for the broken tow, the axial stress along the local fiber orientation is $\sigma^{(k,0)}_{11}(x)$. Let $\delta u^{(k,0)}_{11}(x)$ be the relative displacement between the broken tow and its adjacent neighbors, namely, $\delta u^{(k,0)}_{11}(x)$ is the extra, perturbation displacement caused by the tow break. Initially when there is no broken tow, $\delta u^{(k,0)}_{11}(x)=0$, and all the tows have a common axial strain and uniform longitudinal stress $\sigma^{(k)}_{11,\infty}$. 

The relation between $\delta u^{(k,0)}_{11}$ and $\sigma^{(k,0)}_{11}$ is thus:
\begin{equation}\label{eq:relation_ux_sigmax}
\sigma^{(k,0)}_{11}(x) = \dfrac{d}{dx}\left[\delta u^{(k,0)}_{11}(x)\right] \cdot E_l +  \sigma^{(k)}_{11,\infty}
\end{equation}
where the contribution on $\sigma_{11}$ from the other strain components are neglected in the stress-strain relation. To account for the essential in-situ effects, transverse and through-thickness stresses need to be included because they are correlated with sliding/clamping constraints that will further affect the debonding and stress recovery lengths.

Take a small differentiable tow element of length $dx$, the equilibrium of forces acting on such an element requires that:
\begin{equation}\label{eq:tow_equilibrium_dx}
\frac{d}{dx}\left[ \sigma^{(k,0)}_{11}(x) \right] \cdot (wh) = 2 \cdot \uptau_{\lra}(x) \cdot h + 2 \cdot \uptau_{\uda}(x) \cdot w 
\end{equation} 
where $\uptau_{\lra}$ and $\uptau_{\uda}$ are the intra-ply and inter-ply shear tractions along the interfaces between the broken tow and its surrounding neighbors, respectively.

Since the definition of $\delta u^{(k,0)}_{11}(x)$ is the perturbation displacement of the broken tow, a straightforward way to define interfacial shear strains around the broken tow is:
\begin{equation}\label{eq:interface_shear_strain}
\gamma^{(k,0)}_{\lra}(x) = \frac{\delta u^{(k,0)}_{11}(x)}{t_{\lra}} \, , \qquad 
\gamma^{(k,0)}_{\uda}(x) = \frac{\delta u^{(k,0)}_{11}(x)}{t_{\uda}}
\end{equation}

Let $G_{\lra}$ and $G_{\uda}$ be the shear modulus of corresponding interfaces, whose magnitude should lie within that of the shear modulus of pure resin or the transverse stiffness of the tows, $E_t$.  Considering that the thickness of interfaces are relatively small, thereby define, respectively:
\begin{equation}\label{eq:penaulty_stiffness}
K_{s,\lra} = \frac{G_{\lra}}{t_{\lra}} \;, \qquad K_{s,\uda} = \frac{G_{\uda}}{t_{\uda}} 
\end{equation}
as the effective shear stiffness of the intra-ply and inter-ply interfaces in pure Mode-II or Mode-III shear deformations. This gives a direct relation between the relative displacement and the interfacial shear stresses as:
\begin{align}\label{eq:interface_shear_Ks*u}
\uptau^{(k,0)}_{\lra}(x) &= \gamma_{\lra}(x) \cdot G_{\lra} = \frac{\delta u^{(k,0)}_{11}(x)}{t_{\lra}} = K_{s,\lra} \cdot \delta u^{(k,0)}_{11}(x) \\
\uptau^{(k,0)}_{\uda}(x) &= K_{s, \uda} \cdot \delta u^{(k,0)}_{11}(x)
\end{align}
which has the similar form as the commonly used traction-separation rules in cohesive zone modeling (CZM) techniques.

Sine the interfacial shear stresses depend on status of the interface (still bonded or has failed), whereas the elastic traction-separation response (Eq.\ref{eq:interface_shear_Ks*u}) is only valid for bonded domains, it remains to define the frictional sliding forces in the debonded interfaces. Let $d_{\lra}(x)$, $d_{\uda}(x)$ be the damage variables ($d \in \interval{0}{1}$) of intra-ply and inter-ply interfaces, respectively, with $d=0$ indicating that an interface is intact and $d=1$ means complete interface debond. This kind of application of damage variable is also commonly used in CZM techniques in FEA where the interface penalty stiffness of a cohesive element is typically degraded by a multiplier depending on the damage variables (e.g., $1-d$) to account for interfacial damage and failure status.

\begin{figure}[!htb]
    \centering
    \includegraphics[width=0.5\textwidth]{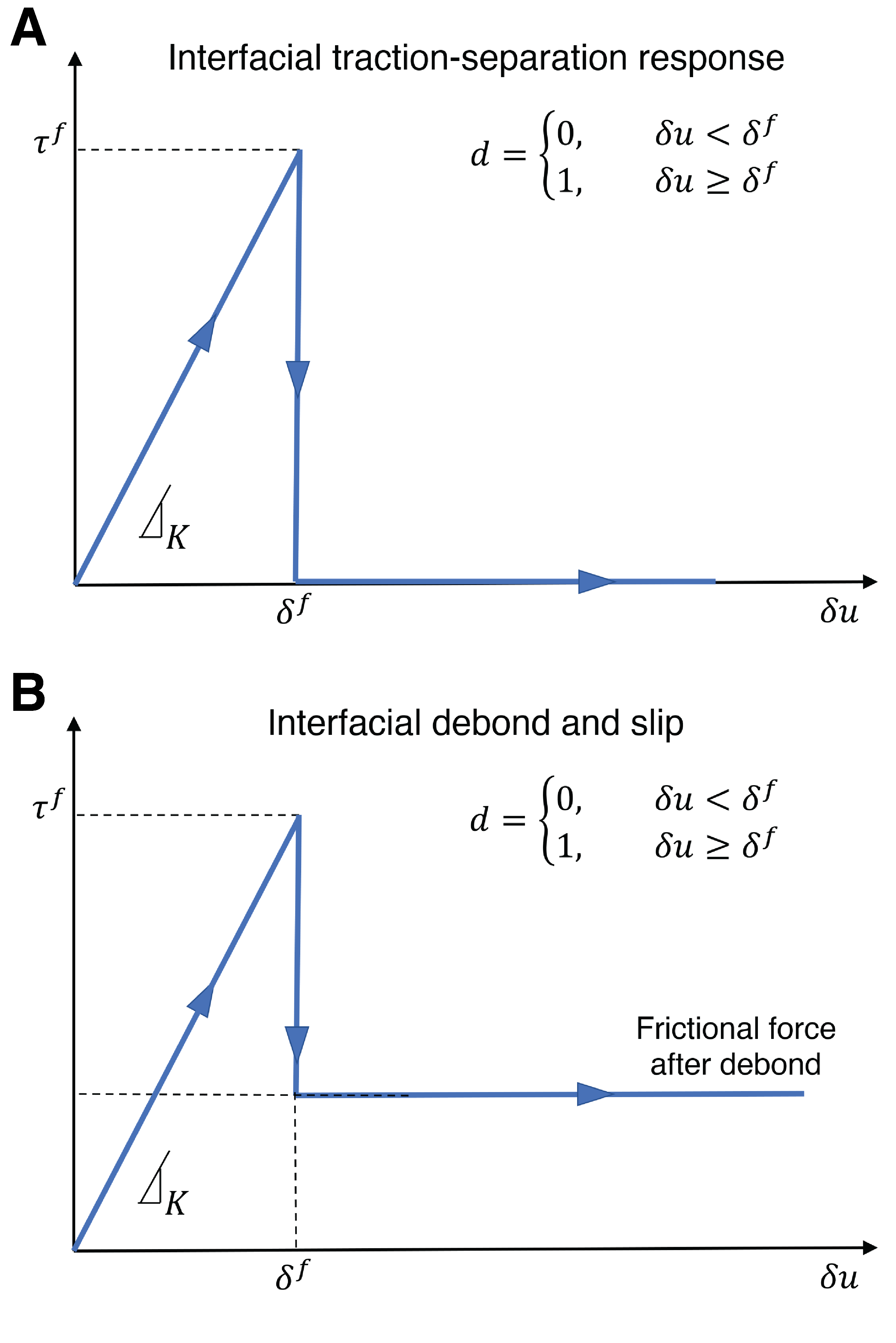}
    \caption{Schematic plot of interfacial behavior in pure shear modes (Mode-II, Mode-III), where $d$ is the effective damage variable of the interface, $K$ is the penalty stiffness, $\delta^f, \tau^f$ are the failure (debonding) separation and traction, respectively. (A) Typical traction-separation response, where the interface debonding occurs upon reaching the failure criterion. (B) If a debonded interfacial region is still in contact and  under negative normal strain, pressure-dependent friction may exist in the interface.}
    \label{fig:debond-and-slip}
\end{figure}

As shown in Fig.\ref{fig:debond-and-slip}, assume that the damage evolution of the interface is rapid, such that the damage variable reaches one instantly when the failure criterion has been reached. Namely, once the interfacial relative displacement has reached a critical value, it breaks instantly. After debonding, the interface may still carry frictional forces depending on the contact pressure between the debonded interface. For pure shear modes (Mode-II and III), assume that a frictional force may exist after debonding (Fig.\ref{fig:debond-and-slip}B), where $\tau^f_s$ and $\tau^f_n$ are the interface strengths in shear and normal modes, respectively, while $\delta^f_s$ and $\delta^f_n$ are the corresponding separation to cause interfacial debond, respectively.

This way, the shear stresses (in both the bonded and debonded interfaces) between the broken tow and its neighbors are expressed in piecewise forms based on the status of the interfaces:
\begin{equation}\label{eq:interface_shear_stresses}
\uptau^{(k,0)}_{\lra}(x) = 
\left\{\begin{array}{lr}
    \delta u^{(k,0)}_{11}(x) \cdot K_{s,\lra} \:,                   & d_{\lra}=0 \vpb             \\
    \mu_{\lra} \cdot \langle - \sigma^{(k)}_{22,\infty} \rangle \:, & d_{\lra}=1 \vphantom{\big|}
\end{array}\right.   \:, \qquad
\uptau^{(k,0)}_{\uda}(x) = 
\left\{\begin{array}{lr}
    \delta u^{(k,0)}_{11}(x) \cdot K_{s,\uda} \:,                 & d_{\lra}=0 \vpb \\
    \mu_{\uda} \cdot \langle -\bar{\sigma}^{(k)}_{33} \rangle \:, & d_{\lra}=1 \vpb
\end{array}\right.
\end{equation}
where $\langle \, \rangle$ denotes the Macaulay bracket:
\begin{equation}
    \langle x \rangle=\max(0, x) =
    \begin{cases}
        x \;, & x \geq 0 \\
        0 \;, & x < 0
    \end{cases}
\end{equation}
while $\mu_{\lra}$ and $\mu_{\uda}$ are the corresponding friction coefficients. 

\subsection{Governing equations and boundary conditions}
Since interfacial shear tractions depends on the interface status (Eq.\ref{eq:interface_shear_stresses}), specific expressions of the force equilibrium (Eq.\ref{eq:tow_equilibrium_dx}) also have to be written in a piecewise form. Thus it is necessary to determine the inequality/equality relation between the debonding lengths, $L_{\lra}$ and $L_{\uda}$, to obtain the domain and expression of differential equations, note that the debonding lengths are also unknown values to be solved.

The force equilibrium of the broken tow (Eq.\ref{eq:tow_equilibrium_dx}) and the physical meaning of perturbation displacement (Eq.\ref{eq:relation_ux_sigmax}) indicates that $\delta u^{(k,0)}_{11}$ is a non-negative, non-increasing function whose magnitude vanishes as $x \to + \infty$. If the shear tractions are small, the stress recover length would be long, and $\delta u^{(k,0)}_{11}$ decrease gradually; if the shear tractions are large, stresses caused by the broken tow would recover to far-field values in a shorter distance.
 
As shown in Fig.\ref{fig:debond-and-slip}, since debonding will only occur upon reaching the interface strength, and all interfaces in the analytical analysis are assumed in pure shear mode, the resultant boundary condition requires that the elastic shear stresses reach the corresponding strengths at the 'debonding tips', the expressions are:
\begin{align}\label{eq:BC_strength_at_debond_tip}
\uptau^{(k,0)}_{\lra}(x) \bigg|_{x = L^{+}_{\lra}} &= \delta u^{(k,0)}_{11}(L^{+}_{\lra}) \cdot K_{s,\lra} = \tau^{f}_{s,\lra} \\
\uptau^{(k,0)}_{\uda}(x) \bigg|_{x = L^{+}_{\uda}} &= \delta u^{(k,0)}_{11}(L^{+}_{\uda}) \cdot K_{s,\uda} = \tau^{f}_{s,\uda}
\end{align}  
where $\tau^{f}_{s,\lra}$ and $\tau^{f}_{s,\uda}$ are the elastic strengths of intra-ply and inter-ply interfaces in pure shear mode, respectively. The superscript $'+'$ on $L_{\lra}$ and $L_{\uda}$ is used to show that function/value need to be evaluated at the bonded domain where damage variables are zero.

Therefore, based on the non-decreasing feature of $\delta u^{(k,0)}_{11}$ and the conditions given in Eq.\ref{eq:BC_strength_at_debond_tip}, it is observed that the relation between $L_{\lra}$ and $L_{\uda}$ depends on the interfacial pure-mode shear strengths and corresponding shear stiffness, which can be categorized into three cases:
\begin{equation}\label{eq:condition_for_3_cases}
    \begin{cases}
        \text{Case 1:} \quad L_{\lra} > L_{\uda}  \:, & \text{if} \quad \tau^f_{s,\lra}/K_{s,\lra} < \tau^f_{s,\uda}/ K_{s,\uda} \vpb \\
        \text{Case 2:} \quad L_{\lra} < L_{\uda}  \:, & \text{if} \quad \tau^f_{s,\lra}/K_{s,\lra} > \tau^f_{s,\uda}/ K_{s,\uda} \vpb \\
        \text{Case 3:} \quad L_{\lra} = L_{\uda}  \:, & \text{if} \quad \tau^f_{s,\lra}/K_{s,\lra} = \tau^f_{s,\uda}/ K_{s,\uda} \vpb 
    \end{cases}
\end{equation}

Now the piecewise governing equations can be written for all cases. Take Case 1 as example, the intervals are $\rinterval{0}{L_{\uda}}$, $\rinterval{L_{\uda}}{L_{\lra}}$, $\rinterval{L_{\lra}}{\infty}$, and the corresponding governing equations are: 
\begin{subnumcases}{ E_l \cdot w h \cdot \dfrac{d^2}{dx^2} \delta u^{(k,0)}_{11} (x)=}
    2h\cdot \mu_{\lra} \cdot \langle {-}\sigma^{(k)}_{22,\infty} \rangle + 2w \cdot \mu_{\uda} \cdot \langle {-}\bar{\sigma}^{(k)}_{33} \rangle \:,  & $ x \in \rinterval{0}{L_{\uda}} \vpb $
    \\
    2h \cdot \mu_{\lra} \cdot \langle {-}\sigma^{(k)}_{22,\infty} \rangle +  2w \cdot K_{s,\uda} \cdot \delta u^{(k,0)}_{11}(x) \:, & $ x \in \rinterval{L_{\uda}}{L_{\lra}} \vpb $
    \\
    2h \cdot K_{s,\lra} \cdot \delta u^{(k,0)}_{11}(x) +  2w \cdot K_{s,\uda} \cdot \delta u^{(k,0)}_{11}(x) \:,  & $ x \in \rinterval{L_{\lra}}{+\infty} \vpb $ 
\end{subnumcases}
which contain three second-order differential equations. Now define the following constants:
\begin{equation}\label{eq:A1_to_A5_definition}
\begin{aligned}
    & A_1 = E_l (w \cdot h), \quad A_2 = 2h\cdot \mu_{\lra} \cdot \langle - \sigma^{(k)}_{22,\infty} \rangle, \quad
    A_3 = 2w\cdot \mu_{\uda} \cdot \langle -\bar{\sigma}^{(k)}_{33} \rangle, \quad \\
    & A_4 = 2h\cdot K_{s,\lra}, \quad A_5 = 2w \cdot K_{s,\uda}
\end{aligned}	
\end{equation}
which are correlated with material properties and in-situ stresses, and are all non-negative constants given the corresponding physical meanings. Then by setting $\delta u \equiv \delta u^{(k,0)}_{11} (x)$ for simplicity, the governing equations for Case 1 can be expressed as:
\begin{subnumcases}{\dfrac{d^2}{dx^2} \delta u =}
    \frac{1}{A_1} (A_2 + A_3),  & $ x \in \rinterval{0}{L_{\uda}} \vpb $ \label{eq:GE.Case1.interv1} \\
    \frac{1}{A_1} (A2 + A5 \cdot \delta u) \:, & $ x \in \rinterval{L_{\uda}}{L_{\lra}} \vpb $ \label{eq:GE.Case1.interv2}
    \\
    \frac{A4 + A5}{A_1} \cdot \delta u,  & $ x \in \rinterval{L_{\lra}}{+\infty} \vpb $ \label{eq:GE.Case1.interv3}
\end{subnumcases}

There are eight undetermined values in the governing equations, two of which are the unknown debonding lengths that will also determine the intervals of piecewise function. There are exactly eight constraints to solve for these unknown values:
\begin{enumerate}[label=(\alph*), nosep]
    \item Axial stress (in fiber orientation) of the broken tow is zero at plane of tow break ($x=0$).
    \item Axial stress recovers to the far-field level after a sufficiently large distance ($x \to +\infty$).
    \item Axial stress is continuous at domain-dividing point $x=L_{\lra}$ of the piecewise function.
    \item Axial stress is continuous at domain-dividing point $x=L_{\uda}$ of the piecewise function.
    \item Perturbation displacement is continuous at $x=L_{\lra}$.
    \item Perturbation displacement is continuous at $x=L_{\uda}$.
    \item Intra-ply interfacial shear stress reaches debonding strength $\tau^{f}_{s,\uda}$ at 'debonding tip' $L^{+}_{\uda}$.
    \item Inter-ply interfacial shear stress reaches debonding strength $\tau^{f}_{s,\lra}$ at 'debonding tip' $L^{+}_{\lra}$.
\end{enumerate}

The expressions of these eight constraints are:\\
\begin{subequations}\label{eq:Case1_BCs}
\noindent\begin{minipage}{.45\linewidth}
\begin{align}
    \text{B.C. for Case 1}: \qquad              & \nonumber                                \\
    \sigma^{(k,0)}_{11}(0)                      & = 0 \vpb                                 \\
    \lim_{x \to +\infty} \sigma^{(k,0)}_{11}(x) & = \sigma^{(k)}_{11,\infty}  \vpb         \\
    \sigma^{(k,0)}_{11}(L^{-}_{\uda})           & = \sigma^{(k,0)}_{11}(L^{+}_{\uda}) \vpb \\
    \sigma^{(k,0)}_{11}(L^{-}_{\lra})           & = \sigma^{(k,0)}_{11}(L^{+}_{\lra}) \vpb
\end{align}
\end{minipage}
\begin{minipage}{.45\linewidth}
\begin{align}
    \text{}                               & \nonumber                                    \\
    \delta u^{(k,0)}_{11}( L^{-}_{\uda} ) & = \delta u^{(k,0)}_{11}( L^{+}_{\uda} ) \vpb \\
    \delta u^{(k,0)}_{11}( L^{-}_{\lra} ) & = \delta u^{(k,0)}_{11}( L^{+}_{\uda} ) \vpb \\
    \delta u^{(k,0)}_{11}( L^{+}_{\uda} ) & = \tau^{f}_{s,\uda} / K_{s,\uda} \vpb        \\
    \delta u^{(k,0)}_{11}( L^{+}_{\lra} ) & = \tau^{f}_{s,\lra} / K_{s,\lra} \vpb
\end{align}
\end{minipage}
\end{subequations}\\
where the superscript $+$ or $-$ on domain-dividing points ($L_{\lra}$ and $L_{\uda}$) indicate the positive or negative side of the interval to evaluate the piecewise function. For example, $x=L^{+}_{\lra}$ means to evaluate Eq.\ref{eq:GE.Case1.interv3} at point $L_{\lra}$, whereas $x=L^{-}_{\lra}$ means to evaluate Eq.\ref{eq:GE.Case1.interv2} at point $L_{\lra}$.

\subsection{Solutions of the shear-lag models}\label{subsec:shearlag-solution}

To solve for all eight unknowns, it is found that a polynomial equation of the debonding lengths have to be obtained first, then the remaining six unknowns can be expressed by either one of the debonding lengths. For Case 3, the total number of unknowns reduces to six but the solution procedure is similar.

Stay with Case 1 as an example, the general solution takes form:
\begin{equation}\label{eq:Case_1_general_solution}
	\delta u^{(k,0)}_{11}(x)=
	\begin{cases}
		\dfrac{A_2 + A_3}{2 A_1} \cdot x^2 - \dfrac{\sigma^{(k)}_{11,\infty}}{E_l} \cdot x + C_2 \:, &   x \in \rinterval{0}{L_{\uda}} \vpb \\
		-A_2/A_5 + C_3 \cdot \exp \left(\sqrt{A_5/A_1}\cdot x \right) + C_4 \cdot \exp \left(-\sqrt{A_5/A_1}\cdot x \right) \;, &  x \in  \rinterval{L_{\uda}}{L_{\lra}} \vpb \\
		C_6 \cdot \exp \left(-\sqrt{(A_4 + A_5)/A_1 }\cdot x \right) \;, &  x \in \rinterval{L_{\lra}}{+\infty} \vpb
	\end{cases}
\end{equation}
The solutions for debonding lengths are:
\begin{equation}\label{eq:Case1_debond_lengths}
\begin{aligned}
    L_{\uda} & = -\dfrac{1}{2A_8} \left( A_9 + \sqrt{A_9^2- 4 A_8 A_{10}}  \right) >0 \,,\vpb \\
    L_{\Delta}  & \equiv L_{\uda} - L_{\lra} = \frac{1}{2} \sqrt{A_1/A_5} \cdot \ln \left(\dfrac{A_6}{A_7} \cdot \dfrac{A_1 B_3 + B_1 L_{\uda}}{A_1 B_2 - B_1 L_{\uda}} \right) < 0 \,,\vpb \\
    L_{\lra} & =  L_{\uda} - L_{\Delta} > L_{\uda} \vpb
\end{aligned}
\end{equation}
where $L_{\Delta}$ is a term showing the difference between two debonding lengths (negative in Case 1), and the remaining four undetermined constants can be expressed by the two debonding lengths:
\begin{equation}\label{eq:Case1_constants_Ci}
\begin{aligned}
& C_2 = -\dfrac{A_2}{A_5} +\dfrac{1}{2A_5 K_{s,\lra}} \left[ A_6\exp\left(-\sqrt{\frac{A_5}{A_1}} \cdot L_{\Delta} \right) + A_7 \exp\left( \sqrt{\frac{A_5}{A_1}} \cdot L_{\Delta} \right)  \right] - \dfrac{A_2 + A_3}{2A_1}\cdot L_{\uda}^2 + \dfrac{\sigma^{(k)}_{11,\infty}}{E_l}\cdot L_{\uda} \,, \vpb \\
& C_3 = \dfrac{A_7}{2A_5 K_{s,\lra}} \exp\left(-\sqrt{A_5/A_1} \cdot L_{\lra} \right) \,, \quad  
C_4 = \dfrac{A_6}{2A_5 K_{s,\lra}} \exp \left( \sqrt{A_5/A_1} \cdot L_{\lra} \right) \,, \vpb \\ 
& C_6 = \dfrac{\tau^f_{s,\lra}}{K_{s,\lra}} \exp \left( \sqrt{(A_4 + A_5)/A_1 }\cdot L_{\lra} \right)
\end{aligned}
\end{equation}
where both Eq.\ref{eq:Case1_debond_lengths} and Eq.\ref{eq:Case1_constants_Ci} contain eight extra constants ($A_6$ to $A_{10}$, $B_1$ to $B_3$), which are dependent values of $A_1$ to $A_5$ and the far-field stress $\sigma^{(k)}_{11,\infty}$. Detailed expressions for these constants and the essential steps to obtain solutions are discussed in appendices.

To strictly ensure uniqueness and physical validity of the shear-lag analysis, there are two required conditions on stress states. The first involves the far-field tensile stress level at which the tow breaks:
\begin{subequations}\label{eq:Case1_constraints_on_stress_states}
\begin{equation}\label{eq:min_vlaue_for_siginfy}
	\left(\sigma^{(k)}_{11,\infty} \right)^2 > \left( \sigma^{*}_{11,\infty} \right)^2 \equiv 
    \frac{A_2 E_l^2}{A_1} \left( \frac{A_2}{A_5} + 2\frac{\tau^f_{s,\uda} }{K_{s,\uda}} \right) 
    - \frac{E_l^2}{A_1} \left[ \frac{A_6 A_7}{A_5 K_{s,\lra}^2} - A_5\left( \frac{\tau^f_{s,\uda} }{K_{s,\uda}} \right)^2 \right]
\end{equation}
where the defined term, $\sigma^{*}_{11,\infty}$, is the minimum threshold value of far-field stress. 
The second condition involves the frictional sliding forces in the debonded/damaged interfaces:
\begin{equation}\label{eq:A2+A3>0}
	2h\cdot \mu_{\lra} \cdot \langle - \sigma^{(k)}_{22,\infty} \rangle + 2w\cdot \mu_{\uda} \cdot \langle -\bar{\sigma}^{(k)}_{33} \rangle = A_2 + A_3 > 0
\end{equation}
\end{subequations}

The physical meaning of the first condition (Eq.\ref{eq:min_vlaue_for_siginfy}) is: if the far-field stress level at which the tow breaks is relatively low, then only the interfacial elastic shear stresses are adequate to counterbalance and recover the axial tensile stress lost by the broken tow, such that the 'debond-and-slip' interfacial failure mode would not take place. The physical meaning of the second condition is: there must be at least some sort of sliding (frictional) forces in the debonded interfaces, so that the broken tow will reach equilibrium after a sufficiently large distance. Otherwise, assume that the far-field stress is large enough to cause interface damage/debonding (Eq.\ref{eq:min_vlaue_for_siginfy} is satisfied) but no frictional forces are accumulated in the debonded interfaces ($A_2+A_3=0$), then the interfacial debonding will propagate continuously such that the damage evolution would become unstable, and no static equilibrium state would be reached. Mathematically, these two conditions will ensure that the solutions of debonding lengths are real and positive values. Derivation of Eq.\ref{eq:Case1_constraints_on_stress_states} is shown in \ref{sec:appendix-solution-procedure}. It is also verified that the solution is uniquely determined when these two conditions are met, the proof is presented in \ref{sec:appendix-proof-of-uniqueness}.

\subsection{A discussion on the interface debonding phenomena}

In the above section, we explain the physical and mathematical meanings of the two required conditions (Eq.\ref{eq:Case1_constraints_on_stress_states}) to ensure the validity of shear-lag analysis and the uniqueness of solutions. 
It is then readily seen that, given the typical dimension and modulus of prepreg tows, the required minimum threshold value of far-field stress ($\sigma^{*}_{11,\infty}$ in Eq.\ref{eq:min_vlaue_for_siginfy}) is approximately two orders of magnitude lower than the typical ultimate tensile stress of tows. This implies that interface debonding is essentially a certain damage mode when a tensile tow break occurs. Namely, the released energy and perturbation forces caused by the broken tow is sufficiently large to trigger debonding along the resin-rich interfaces around the broken tow.
 
This phenomena can also be explained by fracture-mechanics based studies involving a crack at (or approaching) the interface between two continuum domains \cite{He.Hutchinson1989, Gupta.etal1992, Tullock.etal1994, He.etal1994, Parmigiani.Thouless2006, Martin.etal2008}. In such scenarios, the crack may penetrate into the neighboring domain (across interface) or deflect and turn into the bi-material interface. And there is a 'competition' between crack deflection and penetration when determining the path and mode of crack propagation.

A conclusion particularly useful for this study is that when the elastic properties of two continuum domains are identical, crack deflection is almost surely to occur (instead of penetration) if the strength or fracture toughness of the continuum material is more than around four times the mixed-mode toughness of the interface between two domains \cite{He.Hutchinson1989, He.etal1994}. This circumstance applies here because the longitudinal stiffness and fracture toughness of composite tows/strands are typically at least ten times higher than the properties of epoxy resins. Nevertheless, this does not mean that the adjacent tows will not break: tensile failure in neighboring tows (crack penetration) can definitely occur at certain locations if the overall stress level is high; also, interface cracks can possibly turn and penetrate into tows at some 'weak spots' caused by imperfections or flaws along the tow edges. As mentioned, studies involving multiple tow breaks are not in the focus of this paper, instead, it should be performed once a study of a single tow break is established.

\subsection{Summary}
In this section, a shear-lag based analysis of an in-situ tow break in a representative uni-directional filament-wound composite is presented. By considering different inter- and intra-ply debonding lengths, the governing equations are categorized accordingly into different domains, whereby the solutions of axial stress and perturbation displacement of the broken tow are expressed in a piecewise form. Besides, essential variables such as far-field stress states, interfacial shear strengths and frictional coefficients, are included in the analysis to justify the consideration of in-situ effects. In addition, to ensure that the shear-lag analysis is fundamentally meaningful, two required conditions on the in-situ stress states are also determined. The physical meanings of these two conditions are explained, and it is proved that the solutions are uniquely determined when they are satisfied. The general solution for the other two cases are given in \ref{sec:appendix-sol-for-case2} and \ref{sec:appendix-sol-for-case3}.

\section{Stress concentration profile around an in-situ tow break}\label{sec:quantifySCF}

Following the shear-lag analysis, it remains an essential task to evaluate the overload profile and stress concentration factor (SCF) around locations of interest. Due to the complexity of such a problem, closed form analytical solutions of the stress fields are arguably impractical, and yet it is still a challenging task to develop reasonable approximations. 

In this study, we apply a novel approach that combines classic solid mechanics solutions with the shear-lag analysis results from the previous section. Specifically, the idea is to: 1) firstly treat the interfacial shear stresses as 'loads' that generate/exert overload onto the neighboring domains, by considering the fact that interfacial shear stresses are the main 'load transferor' that redistribute the overload to its surrounding intact neighbors; and then 2) predict the stress concentration field by accounting for the interfacial shear tractions based on their magnitude and locations. Detailed steps and derivations are demonstrated in this section.

\subsection{Stress field due to a tangential point load on the surface of a transversely-isotropic half space}\label{subsec:point-load-solution}

The resultant stress field by tangential point load applied on the surface of a transversely-isotropic infinite half space is firstly investigated. In particular, the expressions adopted here is based on the derivations by Lin et al. \cite{Lin.etal1991}. An earlier work by Pan and Chou \cite{Pan.Chou1976} using Green's function solutions was also investigated, but it seems that several constants ('$k_i$' in Eq.30 in their study) were either undefined or misprinted (also implicitly mentioned by Kachanov et al. \cite{Kachanov.etal2003}).

\begin{figure}[!htb]
\centering
\includegraphics[width=0.5\textwidth]{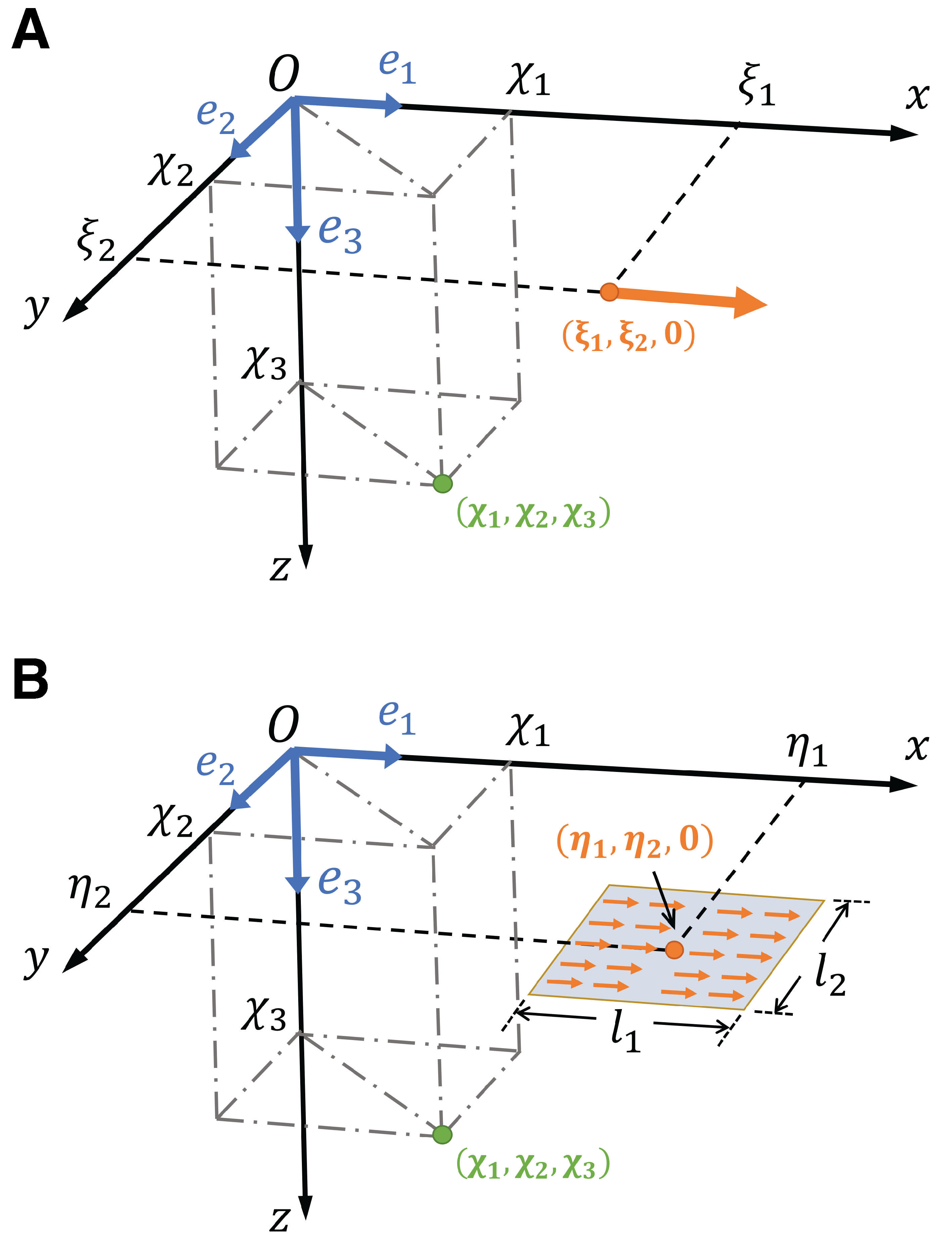}
\caption{Schematic plot of tangential load applied on the surface of an transversely isotropic elastic infinite half space. (A) Tangential point load applied at $(\xi_1, \xi_2, 0)$, with resultant stress component $\sigma_{11}$ at a point of interest $(\chi_1, \chi_2, \chi_3)$ given by Eq.\ref{eq:Omega_11_point_load}. (B) Distributed tangential point load applied over a rectangular region. Eq.\ref{eq:Lambda_11_rec_load} gives the resultant stress component $\sigma_{11}$ at a point $(\chi_1, \chi_2, \chi_3)$, where $(\eta_1, \eta_2, 0)$ is the centroid of the rectangular region, and $l_1, l_2$ are its lengths in $\vec{e_1}, \vec{e_2}$ directions, respectively.}
\label{fig:tangential-load-point-and-rec}
\end{figure}

As illustrated in Fig.\ref{fig:tangential-load-point-and-rec}A, consider a unit tangential point load $\vec{F}= 1 \vec{e_1}$ applied at a point $(\xi_1, \xi_2, 0)$ on the surface of a transversely-isotropic, infinite 3D half space ($z>0$). The resultant axial stress along $\vec{e_1}$ at a point of interest $(\chi_1, \chi_2, \chi_3)$ is:
\begin{equation}\label{eq:Omega_11_point_load}
    \Omega_{11}(\chi_1, \chi_2, \chi_3 \,;\, \xi_1, \xi_2) = 2 c_{66} a_{13} \left( \dfrac{R_2}{D_3^*}\right)_{,12} + 
    \sum_{\alpha=1}^{2} a_{1\alpha} \theta_\alpha \left[ 2 c_{66} \left( \dfrac{R_2}{D_\alpha^*}\right)_{,12} + \dfrac{g_{\alpha} R_1}{D_{\alpha}^3} \right]
\end{equation}
in which the derivative term has form:
\begin{equation}\label{eq:def_of_derivative_(),12}
    \left( \dfrac{R_\alpha}{D_j^*} \right)_{,12} = R_{\alpha} R_1 R_2 \left( \dfrac{1}{ D_j^3 (D_j^*)^2} + \dfrac{2}{D_j^2 (D_j^*)^3} \right) - \dfrac{\delta_{\alpha 2}R_1 + \delta_{\alpha 1}R_2}{D_j (D_j^*)^2}  \:, \qquad \alpha = 1,2, \quad j=1,2,3
\end{equation}
where $\delta_{\alpha \beta}$ is the Kronecker delta operator.
The expressions for all stress components can be found in Refs.\cite{Lin.etal1991, Kachanov.etal2003} (notations may vary). Here only $\sigma_{11}$ is of primary interest. 
The variables in $\Omega_{11}$ that involve coordinates are:
\begin{equation}\label{eq:def_of_coord_vars}
    \begin{aligned}
        & R_\alpha = \chi_{\alpha} - \xi_{\alpha} \:, \quad D_j = \sqrt{R_1^2 + R_2^2 + (\theta_j \chi_j )^2}\:, \quad D_j^* = D_j + \theta_j \chi_j  \:, \qquad \alpha = 1,2, \quad j=1,2,3
    \end{aligned}
\end{equation}
whereas the properties only involve stiffness constants are:
\begin{equation}\label{eq:def_of_constants_PS}
\begin{aligned}
    & \theta_1 = \left[ \dfrac{\sqrt{c_{11} c_{33}} + c_{13} + 2 c_{44} }{4 c_{33}c_{44} (\sqrt{c_{11} c_{33} }-c_{13})^{-1} } \right]^{\frac{1}{2}} 
    + \left[\dfrac{\sqrt{c_{11}c_{33}}-c_{13}-2c_{44} }{4c_{33}c_{44}(\sqrt{c_{11}c_{33}}+c_{13})^{-1} }  \right]^{\frac{1}{2}}  \:, \\
    & \theta_2 = \left[ \dfrac{\sqrt{c_{11} c_{33}} + c_{13} + 2 c_{44} }{4 c_{33}c_{44} (\sqrt{c_{11} c_{33} }-c_{13})^{-1} } \right]^{\frac{1}{2}} 
    - \left[\dfrac{\sqrt{c_{11}c_{33}}-c_{13}-2c_{44} }{4c_{33}c_{44}(\sqrt{c_{11}c_{33}}+c_{13})^{-1} }  \right]^{\frac{1}{2}} \:, \\
    & \theta_3 = \left(\dfrac{c_{66}}{c_{44}}\right)^{\frac{1}{2}} \:, \quad h_1=\dfrac{c_{11}-c_{44}\theta_1^2}{c_{13}+c_{44}}\:, \quad h_2=\dfrac{c_{11}-c_{44}\theta_2^2}{c_{13} + c_{44}} \:, \\
    & b_1 = \theta_1(c_{13}-c_{33}h_1) \:, \quad b_2 = \theta_2(c_{13}-c_{33}h_2) \:, \\
    & g_1 = c_{11}-c_{13}h_1 \:, \quad g_2 = c_{11} - c_{13}h_2 \:, \\
    & a_{11} = \dfrac{b_2}{2\pi (b_1g_2-b_2g_1 )} \:, \quad a_{12} = \dfrac{-b_1}{2\pi (b_1g_2-b_2g_1 )} \:, \quad a_{13} = \dfrac{1}{2\pi c_{44}\theta_3}
\end{aligned}
\end{equation}
in which $c_{ij}$ are components of the compliance matrix of an transversely isotropic material.

The point load solution adopted here requires that $\vec{e_3}$ is normal to the plane of isotropy, i.e., elastic properties in $\vec{e_1}$ and $\vec{e_2}$ directions are identical. Whereas for the purpose of this work, $\vec{e_1}$ is expected to be the longitudinal fiber/tow orientation along which the modulus is much higher than the two transverse directions. However, a particularly important concern in this study is that the tow break is 'in-situ', wherein the following aspects need to be properly treated: 
1) Surfaces of the broken tow are either bonded or remain in contact with neighboring domains, and thus, planes where interfacial shear tractions are applied may not be considered as completely free half surfaces.
2) Sizes of tow break and interface cracks are relatively small compared to typical dimension of the component-scale filament structure, therefore a tow break are not expected to cause large deformations in the surround composite along the transverse directions.
3) Winding patterns may lead to complicated local stacking layups, so that the fiber/tow orientations of adjacent layers may align closely along the transverse direction of the center layer containing the tow break (for instance, a $90^{\circ}$ winding layer above the center $0^{\circ}$ layer). 
Therefore, to address the above issue and ensure compatibility with Eq.\ref{eq:Omega_11_point_load}, the elastic properties in  $\vec{e_1}$ and $\vec{e_2}$ directions are assumed to be identical and comparable to the longitudinal modulus of the tows $E_l$, while the properties in $\vec{e_3}$ direction are equivalent to the transverse modulus of tows $E_t$.

This way, the compliance matrix takes the form:
\begin{equation}\label{eq:def_compliance_Sij}
\begin{bmatrix}
\varepsilon_{11} \\ \varepsilon_{22} \\ \varepsilon_{33} \\ \gamma_{23} \\ \gamma_{31} \\ \gamma_{12} 
\end{bmatrix}
= \begin{bmatrix}
    1/E_l         & -\nu_{ll}/E_l & -\nu_{tl}/E_t & 0        & 0        & 0                   \\
    -\nu_{ll}/E_l & 1/E_l         & \nu_{tl}/E_t  & 0        & 0        & 0                   \\
    -\nu_{lt}/E_l & -\nu_{lt}/E_l & 1/E_t         & 0        & 0        & 0                   \\
    0             & 0             & 0             & 1/G_{tl} & 0        & 0                   \\
    0             & 0             & 0             & 0        & 1/G_{tl} & 0                   \\
    0             & 0             & 0             & 0        & 0        & 2(1+\nu_{ll})/E_{l}
\end{bmatrix} 
\begin{bmatrix}
\sigma_{11} \\ \sigma_{22} \\ \sigma_{33} \\ \sigma_{23} \\ \sigma_{31} \\ \sigma_{12} 
\end{bmatrix}
\end{equation}
where the properties are as defined in Eq.\ref{eq:stress-strain-tow-matrix-form}. Note that the newly introduced Poisson's ratio, $\nu_{ll}$, has to be determined empirically. Briefly, the concept of $\nu_{ll}$ is the in-plane Poisson's ratio of a thin plate stacked with sufficiently large amount of tows in randomly distributed orientations, therefore a reasonable approximation is to take the typical value of balanced woven fabric composite made of materials with comparable modulus. For instance, an approximated value for balanced Carbon/Epoxy woven composites is $\nu_{ll}=0.045$, and $0.13$ for E-Glass \cite{Gay2014} composites, respectively. Nevertheless, the effect of picked values for Poisson's ratios is found to be negligible comparing to other properties. Then upon definition of the compliance matrix, the expressions for the required stiffness constants in Eq.\ref{eq:def_of_constants_PS} can be readily obtained by taking inverse of the compliance matrix in Eq.\ref{eq:def_compliance_Sij}.

\subsection{Stress field due to uniformly distributed tangential load over a rectangular region on the surface of a transversely-isotropic half space}\label{subsec:rec-patch-solution}

For the purpose of this study, a specific interest is the axial stress field (in $\vec{e_1}$ direction) caused by uniformly distributed tangential load over a rectangular region, because this is the 'shape' of debonded domain among which interfacial shear tractions occur. As schematically shown in Fig.\ref{fig:tangential-load-point-and-rec}B, by integrating the point load solution (Eq.\ref{eq:Omega_11_point_load}) over a rectangular region \cite{Lin.etal1991}, and then by applying necessary coordinate transformation, the resultant axial stress caused by tangential load over a rectangular region on the surface of half space can be expresses as:
\begin{multline}\label{eq:Lambda_11_rec_load}
    \Lambda_{11}(\chi_1, \chi_2, \chi_3 \,;\, \eta_1, \eta_2 \,;\, l_1, l_2) = \quad
    \Omega_{11}(\chi_1{+}\eta_1, \chi_2{+}\eta_2, \chi_3 \,;\, {-}\frac{l_1}{2}, {-}\frac{l_2}{2})
    +\Omega_{11}(\chi_1{+}\eta_1, \chi_2{+}\eta_2, \chi_3 \,;\, \frac{l_1}{2}, \frac{l_2}{2}) \\
    -\Omega_{11}(\chi_1{+}\eta_1, \chi_2{+}\eta_2, \chi_3 \,;\, {-}\frac{l_1}{2}, {+}\frac{l_2}{2}) 
    -\Omega_{11}(\chi_1{+}\eta_1, \chi_2{+}\eta_2, \chi_3 \,;\, \frac{l_1}{2}, {-}\frac{l_2}{2})
\end{multline}
where $(\xi_1, \xi_2, 0)$ is the coordinate of centroid of the rectangular region, and $l_1, l_2$ are its lengths in the $\vec{e_1}$ and $\vec{e_2}$ directions, respectively.

\subsection{Critical stress concentration profile along intra-ply neighboring tows}\label{subsec: intra-ply-SCF}
By utilizing the above formulas, overload caused by the broken tow can be estimated by accounting for the contributions from interfacial shear tractions. In particular, the following procedures are used to compute overload profile at a point of interest:
\begin{itemize}
    \item Take each of the eight debonded interfaces around the broken tow, respectively, as the 'surface of transversely isotropic elastic half space'.
    \item Based on the normal directions of all debonded interfaces and the position of the point of interest, apply coordinate transformation, and then compute the resultant overload stress caused by interfacial frictional shear forces in each debonded interface, respectively. 
    \item By combining together the contribution from each debonded interface, the overload stress at any point can be obtained.
\end{itemize}

Note that the solution of stress field ($\Omega_{11}$ in Eq.\ref{eq:Omega_11_point_load}) have singularities at the location of point load. After applying integration over a rectangular region ($\Lambda_{11}$ in Eq.\ref{eq:Lambda_11_rec_load}), logarithmic singularity still exists at the boundaries (corner and edges) of the rectangular region, therefore, direct evaluation based on numerical integration is not valid at those isolated locations. 

Nevertheless, the main interest in this study is the 'critical' stress concentration profile in the plane of tow break ($x=0$), or more specifically, those along the y- and z-axis, because these are the 'plane' and 'paths' where SCF are most prominent. Since the crack opening displacement is positive ($\delta_{cod}>0$), overload profile in the entire tow break plan (including the y- and z-axis) does not have singularities. Besides, since the shear stress in the deformed-but-bonded interfaces are away from tow breakage and also decay rapidly, the contribution from elastic shear stresses in the bonded interfaces are neglected from the quantification of overload profile if the point of interest is in the plane of tow break.

Following the above discussion, the 'critical' stress magnitude in the intra-ply neighboring tows along the y-axis can be evaluated by:
\begin{equation}\label{eq:SCF_yAxis}
\sigma^{(k)}_{11,\text{yaxis}} |_{y>w/2} = \sigma^{(k)}_{11,\infty} 
+ \mu_{\lra} \cdot \langle {-}\sigma^{(k)}_{22,\infty} \rangle \cdot \Lambda^{\Sigma, \text{intra}}_{11,\text{pert},y} 
+ \mu_{\uda} \cdot \langle {-}\bar{\sigma}^{(k)}_{33} \rangle \cdot \Lambda^{\Sigma, \text{inter}}_{11,\text{pert},y}  
\end{equation}
where $\sigma^{(k)}_{11,\infty}$ and $\sigma^{(k)}_{22,\infty}$ are the far-field longitudinal and transverse stress magnitudes in the local coordinate of tows, $\bar{\sigma}^{(k)}_{33}$ is the average through-thickness stress across the $k$th ply (as defined in Eq.\ref{eq:far_field_stress}). 

In Eq.\ref{eq:SCF_yAxis}, $\Lambda^{\Sigma, \text{intra}}_{11,\text{pert}}$ and $\Lambda^{\Sigma, \text{inter}}_{11,\text{pert}}$ are the $'11'$ component of overload (perturbation) stresses along $y$ axis caused by unit shear tractions in the intra-ply and inter-ply interfaces, respectively, which take forms:
\begin{align}\label{eq:SCF_yAxis_Lambdas}
    \begin{split}
        \Lambda^{\Sigma, \text{intra}}_{11,\text{pert},y} = {} &  \Lambda_{11}(0,0,y{-}\frac{w}{2} \,;\, \delta_{cod}{+}\frac{L_{\lra}}{2} , 0 \,;\, L_{\lra}, h )
        - \Lambda_{11}(0,0,y{-}\frac{w}{2} \,;\, {-}\delta_{cod}{-}\frac{L_{\lra}}{2} , 0 \,;\, L_{\lra}, h ) \\
        & + \Lambda_{11}(0,0,y{+}\frac{w}{2} \,;\, \delta_{cod}{+}\frac{L_{\lra}}{2} , 0 \,;\, L_{\lra}, h )
        - \Lambda_{11}(0,0,y{+}\frac{w}{2} \,;\, {-}\delta_{cod}{-}\frac{L_{\lra}}{2} , 0 \,;\, L_{\lra}, h ) 
    \end{split}\\
        \Lambda^{\Sigma, \text{inter}}_{11,\text{pert},y} = {} & 2 \left[ \Lambda_{11}(0,y,\frac{h}{2} \,;\, \delta_{cod}{+}\frac{L_{\uda}}{2} , 0 \,;\, L_{\uda}, w ) - \Lambda_{11}(0,y,\frac{h}{2} \,;\, {-}\delta_{cod}{-}\frac{L_{\uda}}{2} , 0 \,;\, L_{\uda}, w )  \right]
\end{align}	
where the input variables in stress functions ($\Lambda_{11}$ in Eq.\ref{eq:Lambda_11_rec_load}) are replaced with the material and geometrical parameters correlated to the modulus of tows and solved deformations from the shear-lag algorithms. The four components in $\Lambda^{\Sigma, \text{intra}}_{11,\text{pert}}$ refer to the four debonded interface regions in the same ply ($y = \pm w/2$ planes), where coordinate involving $y$ are placed as the third variable due to definition of $\Lambda_{11}$ and coordinate transformation. The two terms in $\Lambda^{\Sigma, \text{inter}}_{11,\text{pert}}$ refer to the two debonded interface regions between the broken tow and its neighboring ply on the top ($z=h/2$) planes, and since the debonded domains in the above and below surfaces of the broken tow (in $z=\pm h/2$ planes) are symmetric about the $z=0$ plane, a coefficient of 2 is applied. Difference in signs of $\Lambda_{11}$ are due to the opposite directions of the shear tractions relative to the positive direction of y-axis.

\subsection{Critical stress concentration profile along inter-ply neighboring tow regions}\label{subsec: inter-ply-SCF}
For the stress redistribution between intra-ply tows as shown above, it is relatively straightforward since tows in the same ply are parallel to each other. In contrast, tow orientations (winding angles) can vary across different plies, and hence quantification of overload profile requires extra consideration. This is an important aspect for typical FWCs because winding patterns not only determine the effective properties of composites, but also affect the crack propagation and damage mechanisms.

As shown schematically in Fig.\ref{fig:csys-and-notation}C, tows in other plies are regarded as 'tow regions' if winding angles are different. Despite that, critical stress concentration is still expected in the plane of broken tow along z-axis. So the idea here is to: 1) compute the 'reference' overload profile for the case of uni-directional tows where all plies are assumed to have the same winding angle as the $k\text{th}$ ply, and then 2) to predict the 'real' overload profile based on basis transformation corresponding to a certain winding angle. 

Therefore, by firstly assuming that all winding layers are locally uni-directional, the stress concentration field (denoted by superscript 'scf,uni') along the z-axis, can be expressed as:
\begin{equation}\label{eq:SCF_zAxis_uni}
	\sigma^{\text{scf,uni}}_{11,\text{zaxis}} |_{z> h/2} = 
	 \mu_{\lra}\cdot \langle {-}\sigma^{(k)}_{22,\infty} \rangle \cdot \Lambda^{\Sigma, \text{intra}}_{11,\text{pert,z}} 
	+ \mu_{\uda}\cdot \langle {-}\bar{\sigma}^{(k)}_{33} \rangle \cdot \Lambda^{\Sigma, \text{inter}}_{11,\text{pert,z}}  
\end{equation}
where $\Lambda^{\Sigma, \text{intra}}_{11,\text{pert,z}}$ and $\Lambda^{\Sigma, \text{inter}}_{11,\text{pert,z}}$ are the $'11'$ component of overload (perturbation) stresses along z-axis caused by unit shear tractions in the intra-ply and inter-ply interfaces around the broken tow, respectively, which take forms:
\begin{align}\label{eq:SCF_zAxis_Lambdas_uni}
    \begin{split}
        \Lambda^{\Sigma, \text{intra}}_{11,\text{pert,z}} = {} & \Lambda_{11}(0,0,z{-}\frac{h}{2} \,;\, +\delta_{cod}{+}\frac{L_{\uda}}{2} , 0 \,;\, L_{\uda}, w ) 
        - \Lambda_{11}(0,0,z{-}\frac{h}{2} \,;\, {-}\delta_{cod}{-}\frac{L_{\uda}}{2} , 0 \,;\, L_{\uda}, w ) \\
        & + \Lambda_{11}(0,0,z{+}\frac{h}{2} \,;\, {+}\delta_{cod}{+}\frac{L_{\uda}}{2} , 0 \,;\, L_{\uda}, w ) 
        - \Lambda_{11}(0,0,z{+}\frac{h}{2} \,;\, {-}\delta_{cod}{-}\frac{L_{\uda}}{2} , 0 \,;\, L_{\uda}, w )
    \end{split}\\
    \Lambda^{\Sigma, \text{inter}}_{11,\text{pert},z} = {} & 2 \left[ \ \; \Lambda_{11}(0,z,\frac{w}{2} \,;\, \delta_{cod}{+}\frac{L_{\lra}}{2} , 0 \,;\, L_{\lra}, h ) -\Lambda_{11}(0,z,\frac{w}{2} \,;\, {-}\delta_{cod}{-}\frac{L_{\lra}}{2} , 0 \,;\, L_{\lra}, w )  \right]
\end{align}

Then for the $i$th ply ($i \neq k$), the stress concentration in the local fiber orientation, denoted by superscript 'scf,local', is approximated by simple change of basis:
\begin{equation}\label{eq:SCF_zAxis_local}
	\sigma^{\text{scf,local}}_{11,\text{zaxis}} |_{z \in [z_i, z_{i+1}] } = \cos\theta^{i} \cdot \sigma^{\text{scf,uni}}_{11,\text{zaxis}} |_{z \in [z_i, z_{i+1}] }
\end{equation}
where $\theta^{i} \in [-90^{\circ}, 90^{\circ}]$ is the valid interval for ply winding angles. It is straight forward to see that for $\pm 90^{\circ}$ plies, the stress concentration in local fiber orientation becomes zero, and reaches maximum when the winding angle is the same as broken tow $0^{\circ}$.

\subsection{Summary}
In this section, by combining: 1) the solution of interface debonding lengths and tow break opening displacement from shear-lag analysis, 2) solutions of resultant stresses caused by tangential loads applied on the surface of a transversely-isotropic linear elastic half space, and 3) superposition concept and coordinate transformations, we derived formulas to quantify critical stress concentration profiles in the plane of tow break. Such an approach requires more calculations than simple empirical formulas, but the discrete cracks and damage patterns are treated in a more 'high-fidelity' way. In the following section, the proposed approach will be compared with a series of finite element studies.

\section{Finite element validation using AGDM}\label{sec:feaValidation}

Numerical simulations to predict tow break opening, debonding lengths, and stress concentration profiles are carried out by implementing an auto-generated geometry-based discrete finite element model (AGDM) recently proposed by Liu and Phoenix \cite{Liu.Phoenix2020}, which has been developed in terms of Abaqus processing commands, and hence is compatible with Abaqus/CAE environment. AGDM realizes accurate coupling between all three types of typical damage modes in laminated composite structures (fiber/yarn break, matrix crack, and delamination) having arbitrary stacking layup angles, and hence it is adopted here to effectively model tow/yarn break, debonding between yarns (matrix crack), and inter-ply debonding (delamination).

Briefly speaking, in order to represent a crack in an explicit way and ensure accurate displacement field across the crack tip, two types of approaches may be implemented: 1) a solid element must be partitioned into sub-domains, or 2) an embedded cohesive element (between solid elements) must deform and fail, to ensure real discontinuity and fracture surfaces in the numerical space. For interaction and coupling of multiple cracks, these discontinuity (fracture surfaces) must have matched mesh at the potential intersecting/deflecting locations \cite{Liu.Phoenix2020}. Unlike typical approaches that rely on bonded mesh to realize matching of mesh, AGDM is created as an assembly of discrete continuum domains and small-thickness interfaces with matching of nodes at potential crack coupling locations, and thereby ensures accurate displacement jump across crack tips and coupling between cracks without requiring priory-generated (bonded) solid elements. This advantage firstly helps to create numerical model for laminates with arbitrary stacking layup angles, and secondly provides the flexibility to use extra fine meshes (for cohesive and/or continuum elements) to capture high-fidelity stress gradients. Details on the algorithms and generation procedures of AGDM can be found in Refs.\cite{Liu.Phoenix2020, liu2020Thesis}.

An earlier numerical study attempting to quantify SCF around a broken tow was also reported by Liu and Phoenix \cite{Liu.Phoenix2019}, but only limited success was obtained in single-layer models (without inter-ply debonding) due to non-matched mesh between inserted cohesive elements at potential crack deflection locations. Consequently, the multi-ply scenarios were not included due to converging issues and artificial stress profiles mostly caused by ambiguous interpolation rules between continuum and cohesive domains. Later on, AGDM was developed after observing the limitations in existing methods, and thereafter is implemented in this study to verify the newly derived analytical approach in a high-fidelity way. 

\subsection{Material properties and numerical inputs}\label{subsec:inputs-for-fea}
The properties of prepreg tows used for this study are based on data sheet of a commercially available, high strength Toray\textsuperscript{\textregistered} T1100g 12k tows. The mechanical, dimensional, and interfacial properties of T1110 layers are listed in Table \ref{tab:t1100g}, where the same nomenclature as in previous sections is used.
The cross-sectional area is approximated based on the nominal void volume fraction and number of filaments, width and height of tow cross section is then obtained by assuming a width-to-height ratio of $3:1$. It should be noted that several parameters (such as the shear modulus of tows $G_{lt}$) are estimated based on reported value of comparable materials. And the main purpose of FEA here is to compare with the prediction from the derived algorithms in Section \ref{sec:quantifySCF} using the same input values.

The interfacial behaviors (i.e., traction-separation rules in bonded interfaces, and pressure-dependent sliding in debonded interfaces, respectively) as explained in Section \ref{subsec:CZM} for intra-ply and inter-ply cohesive elements are realized by user material subroutines (VUMAT).

\renewcommand{\arraystretch}{1.25}
\begin{table}[ht]
\centering
\caption{Mechanical, dimensional, and interface properties of T1100g 12k resin-impregnated tows} \label{tab:t1100g}
\begin{tabular}{c l c c}
    \toprule
    & \makecell{Definition} & Variable & Value \\ 
    \cmidrule{2-4}
    \multirow{6}{*}{\makecell{Mechanical \\ constants}} & Tensile modulus of tow in local longitudinal orientation & $E_l$ & 185.0 GPa \\
    & Tensile modulus of tow in transverse orientations & $E_t$ & 10.0 GPa \\
    & Shear modulus of tow in longitudinal-to-transverse orientation & $G_{lt}$ & 6.5 GPa \\
    & Poisson's ratio of tow in longitudinal-to-transverse orientation & $\nu_{lt}$ & 0.25 \\
    & Poisson's ratio of tow in transverse-to-transverse orientation & $\nu_{tt}$ & 0.40 \\
    & Approximated planar Poisson's ratio of randomly stacked plies & $\nu_{ll}$ & 0.045 \\ \cmidrule{2-4}
    \multirow{3}{*}{ \makecell{Cross-section \\ dimension} } & Representative cross-sectional area & $A_{tow}$ & $0.39$ $\text{mm}^2$ \\ 
    & Representative cross-sectional width & $w$ & $1.09$ mm \\
    & Representative cross-sectional height (along stacking direction) & $h$ & $0.36$ mm \\ \cmidrule{2-4}
    \multirow{4}{*}{\makecell{Intra-ply\\ properties}} & Interface penalty stiffness & $K_{s, \lra}$ & $10^6$ N/$\text{mm}^3$ \\
    & Interface strength in pure normal mode (Mode-I) & $\tau^{f}_{n,\lra}$ & 80.0 MPa \\
    & Interface strength in pure shear modes (Mode-II or -III) & $\tau^{f}_{s,\lra}$ & 100.0 MPa \\
    & Interface frictional coefficient (after debonding) & $\mu_{\lra}$ & 0.30\\ \cmidrule{2-4}
    \multirow{4}{*}{\makecell{Inter-ply\\ properties}} & Interface penalty stiffness & $K_{s, \uda}$ & $10^6$ N/$\text{mm}^3$ \\
    & Interface strength in pure normal mode (Mode-I) & $\tau^{f}_{n,\uda}$ & 80.0 MPa \\
    & Interface strength in pure shear modes (Mode-II or -III) & $\tau^{f}_{s,\uda}$ & 120.0 MPa \\
    & Interface frictional coefficient (after debonding) & $\mu_{\uda}$ & 0.45\\ 
    \bottomrule
    \end{tabular}
\end{table}

\subsection{Model setup and boundary conditions}
The representative FWC laminate is 100mm long along the x-axis (also the fiber orientation), a length found to be sufficiently large such that the longitudinal stress in broken tow recovers to far-field level before reaching boundaries of the model. There are 4 neighboring tows on each side of the broken tow in the center, so the width of the laminate (dimension along the y-axis) is $w \times 9= 9.81mm$. There are in total 13 plies, 6 above and 6 below the $k$th ply, respectively, so the total thickness of laminate is $h \times 13= 4.68mm$. The center (origin) of the tow break plane is placed to overlap with the origin of global Cartesian coordinate system (Fig.\ref{fig:cutView3D-Intra} and Fig.\ref{fig:cutVew3D-Inter}). 

\begin{figure}[!htb]
\centering
\includegraphics[width=0.95\textwidth]{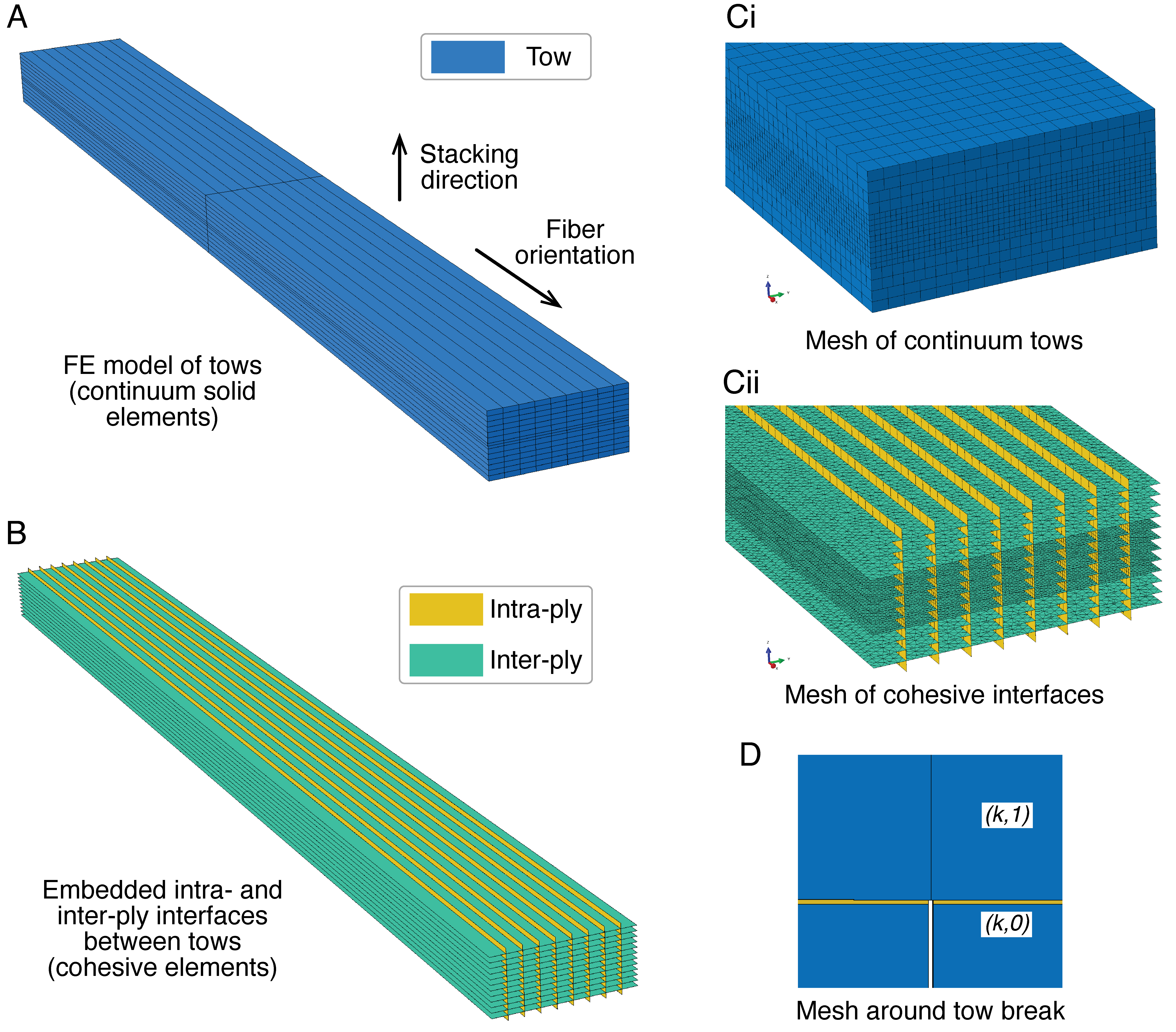}
\caption{Finite element model and mesh. 
(A) Continuum tows (no mesh) with partitions along the $x=0$ and $z=0$ planes to ensure existing of element nodes in these planes (after mesh) for evaluation of critical overload profile. 
(B) Embedded small-thickness interfaces (no mesh) between the continuum tows for potential interfacial debonding and frictions. 
(Ci) Mesh of continuum tows, where the tows in the plies close to the center ply are assigned with smaller element sizes to realize fine stress gradient.
(Cii) Mesh of embedded interface domains (into cohesive elements) between the continuum tows.
(D) Mesh around the broken tow. In this study, breakage (fracture) is accounted for by a discontinuity (removed cohesive element) in the center of the broken tow. The initial gap between the fracture surfaces is $10^{-6}$mm. Matching of mesh is ensured to realize accurate crack deflection between the tow break and both intra-ply and inter-ply interfaces.}
\label{fig:FE-model-snap}
\end{figure}

Fig.\ref{fig:FE-model-snap}A shows the continuum tows in the FE model. Fig.\ref{fig:FE-model-snap}B shows the embedded intra- and inter-ply small-thickness interfaces (to be meshed to cohesive elements) among tows to capture potential interface debonding. Fig.\ref{fig:FE-model-snap}C contains two figures showing the mesh of continuum and cohesive domains, respectively. Since the tow break occurs at the center of the 7th ply, tows in the 5th to 9th plies are all assigned with a smaller element size to ensure that there are at least eight fully-integrated solid elements across a tow's width, and at least two layers of elements across a tow's height. This helps to capture high-fidelity stress gradients around the broken tow, as will be shown in the next subsection.

Since the present study is focused on the equilibrium state whereby any dynamic effect is neglected, the broken tow is represented by an initial slit (discontinuity) with extra small gap ($10^{-6}$mm, see Fig.\ref{fig:FE-model-snap}D). This way, after the model reaches assumed stress states, the resultant debonding lengths, break opening displacement, and surrounding stress fields are equivalent to the static equilibrium state where a tow break would result at the same stress states. More specifically, under the framework of AGDM, one single tow breakage interface domain at the center of the 6th ply is embedded into the center tow. This interface is then removed from the numerical model so that the model initially contains a 'closed' tow break. Then during numerical analysis, as the far-field stresses increases, the tow break surfaces will open up and lead to debonding due to excessive interfacial shear deformations.

The boundary conditions are determined based on the targeted far-field stress magnitudes (Eq.\ref{eq:far_field_stress}). Once these targeted stresses are determined, the corresponding strains are calculated based on the effective properties of the representative laminate (Eq.\ref{eq:def_compliance_Sij}). Then the targeted strains are applied in terms of velocity load on all the element nodes on the outer surfaces of the laminate model. This method of applying boundary conditions is specifically adopted since it is found to alleviate kinetic energy as compared to directly applying stresses. Kinetic energy is monitored throughout the numerical analysis, and ensured to be always less than $1\%$ of the strain energy of the model, to justify that the system can be considered as quasi-static.

Assume that when the tow breaks, the tensile stress along the fiber orientation is $\sigma^{(k)}_{11,\infty}= 1000\text{MPa}$. And the in-situ transverse stresses ($\sigma^{(k)}_{22,\infty}$, $\sigma^{(k)}_{33,\infty}$) are picked from the possible combinations of three values: -50, -100, and -150MPa, resulting in 9 cases in total (Table.\ref{tab:comparison-FEA-theory}). Note that $\bar{\sigma}^{(k)}_{33}$ is in general negative for FWC structures (e.g., COPVs) due to applied internal pressure load. For uni-directional tows in this study, the transverse stress $\sigma^{(k)}_{22,\infty}$ is also considered to be negative here due to Poisson's effect. In reality, a piece of a representative overwrap model may be under bi-axial tension.

\subsection{Examples of FEA results}
\begin{figure}[!htb]
\centering
\includegraphics[width=0.95\textwidth]{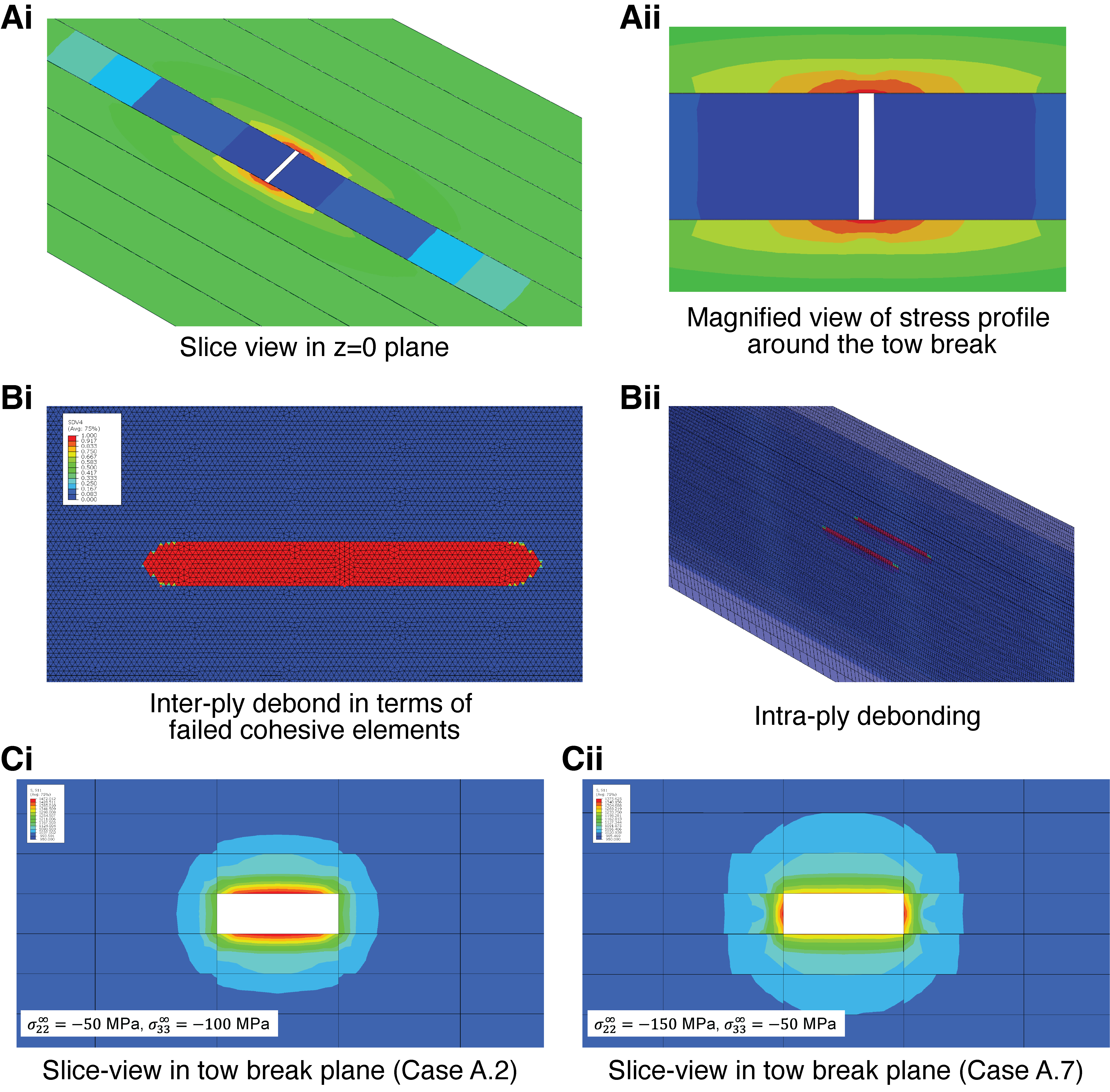}
\caption{Finite element results showing deformation, interface debonding, and stress profile around the broken tow. 
(A) Slice view of the deformation around the tow break at the $z=0$ plane, showing break opening and axial stress profile ($\sigma_{11}$).
(B) Predicted interface debonding in the form of failed cohesive elements, wherein the failed elements (with damage variable equals one) are shown in red, and the intact elements are shown in blue.
(C) Slice views at plane of tow break ($x=0$ plane) with two different in-situ stress states, showing effect of the stress states on the stress concentration profile and factors.}
\label{fig:FE-results-snap}
\end{figure}

Here we briefly show several results on deformation, stress redistribution and concentration, and also interface debonding from FEA. For example, Fig.\ref{fig:FE-results-snap}A shows slice views at the mid-plane ($z=0$) of the model, it can be observed that the stress lost by the broken tow is redistributed to the neighboring units, and the tensile stress gradually recovers to the far-field level. Fig.\ref{fig:FE-results-snap}B shows the predicted interface debonding in terms of failed and damaged cohesive elements, where an element shown in red indicates complete failure (its damage variable reaches one). The debonding lengths can thus be obtained by measuring the length of the failed elements from the plane of tow break to the furthest x-coordinate with failed elements. Fig.\ref{fig:FE-results-snap}C contains two examples showing the slice views at plane of two break, it can be seen that the in-situ stress states affect the stress redistribution mechanisms and subsequently the overload profile and SCFs. 

\subsection{Comparison between numerical and analytical predictions}

To verify the accuracy and robustness of the derived algorithms, we compare the analytical predictions with the numerical results. A specific interest is the overload profile and SCF in the plane of tow break, because SCFs are expected to be most prominent in this plane, and henceforth the maximum errors between FEA and theory would also occur in this plane since stresses will decay from this plane. More specifically, the y-axis and z-axis are expected to be the 'path' along which the intra- and inter-ply stress concentrations are most significant, respectively. 

\begin{figure}[!htb]
\centering
\includegraphics[width=0.75\textwidth]{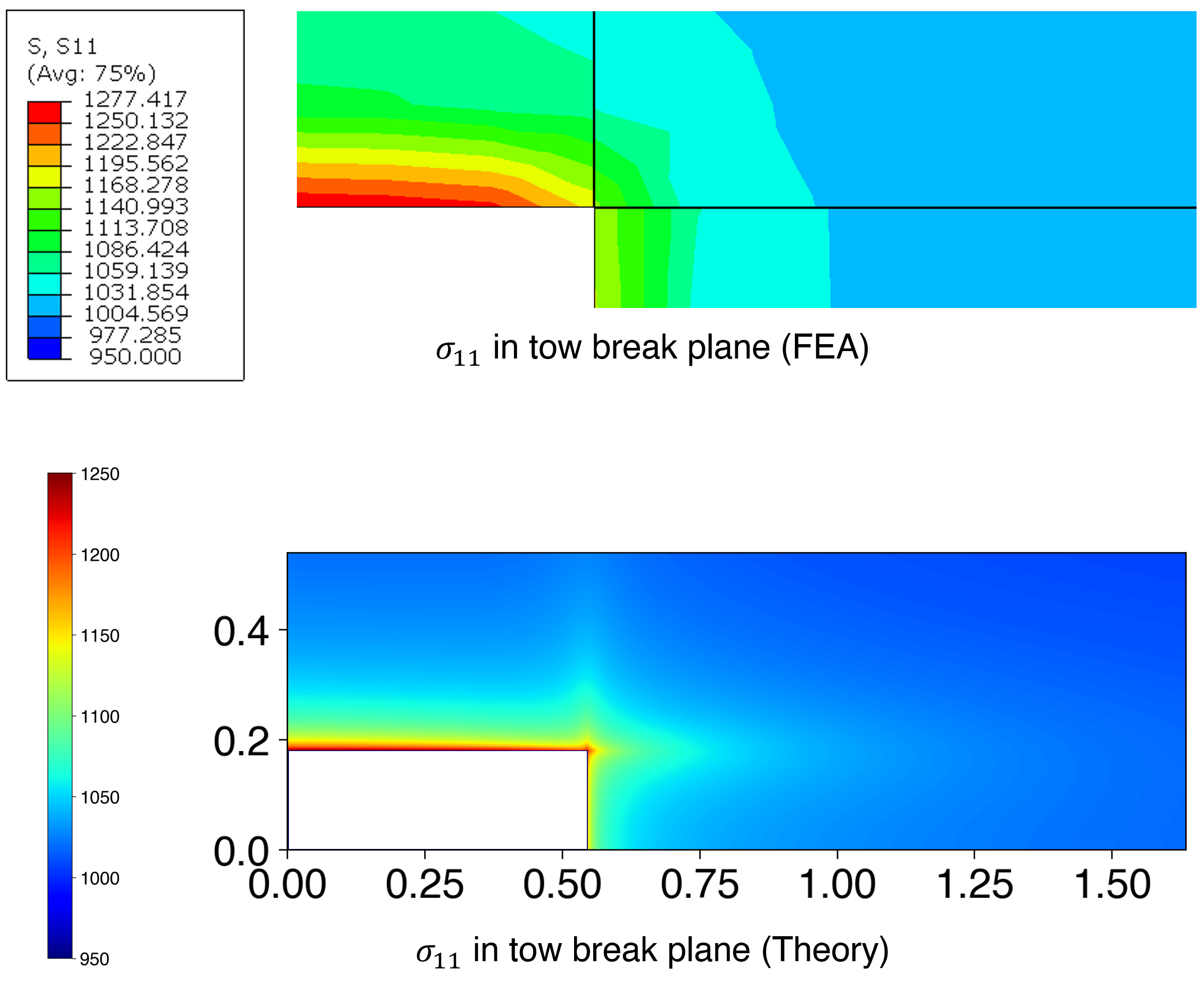}
\caption{Comparison between FEA and theory on the predicted tensile stress in tow break plane, with in-situ stresses $\sigma^{(k)}_{11,\infty}=1000$, $\sigma^{(k)}_{22,\infty}= -50$, and $\bar{\sigma}^{(k)}_{33}=-50$ (Case A.1 in Table.\ref{tab:comparison-FEA-theory}). (A) Stress profile by FEA. (B) Predicted stress profile by analytical approach (Eq.\ref{eq:Lambda_11_rec_load}).}
\label{fig:comparison-sig11-bkPlane}
\end{figure}

Fig.\ref{fig:comparison-sig11-bkPlane} shows a comparison between the predicted tensile stress along fiber orientation ($\sigma_{11}$) in the tow break plane for Case A.1 ($\sigma^{(k)}_{22,\infty}=-50$, $\sigma^{(k)}_{33,\infty}=-50$). The stress profile predicted by theory (Eq.\ref{eq:Lambda_11_rec_load}) agrees reasonable well with the prediction by FEA. Moreover, the maximum stresses (and hence the maximum SCF) also matches closely with each other.

\begin{figure}[!htb]
\centering
\includegraphics[width=0.75\textwidth]{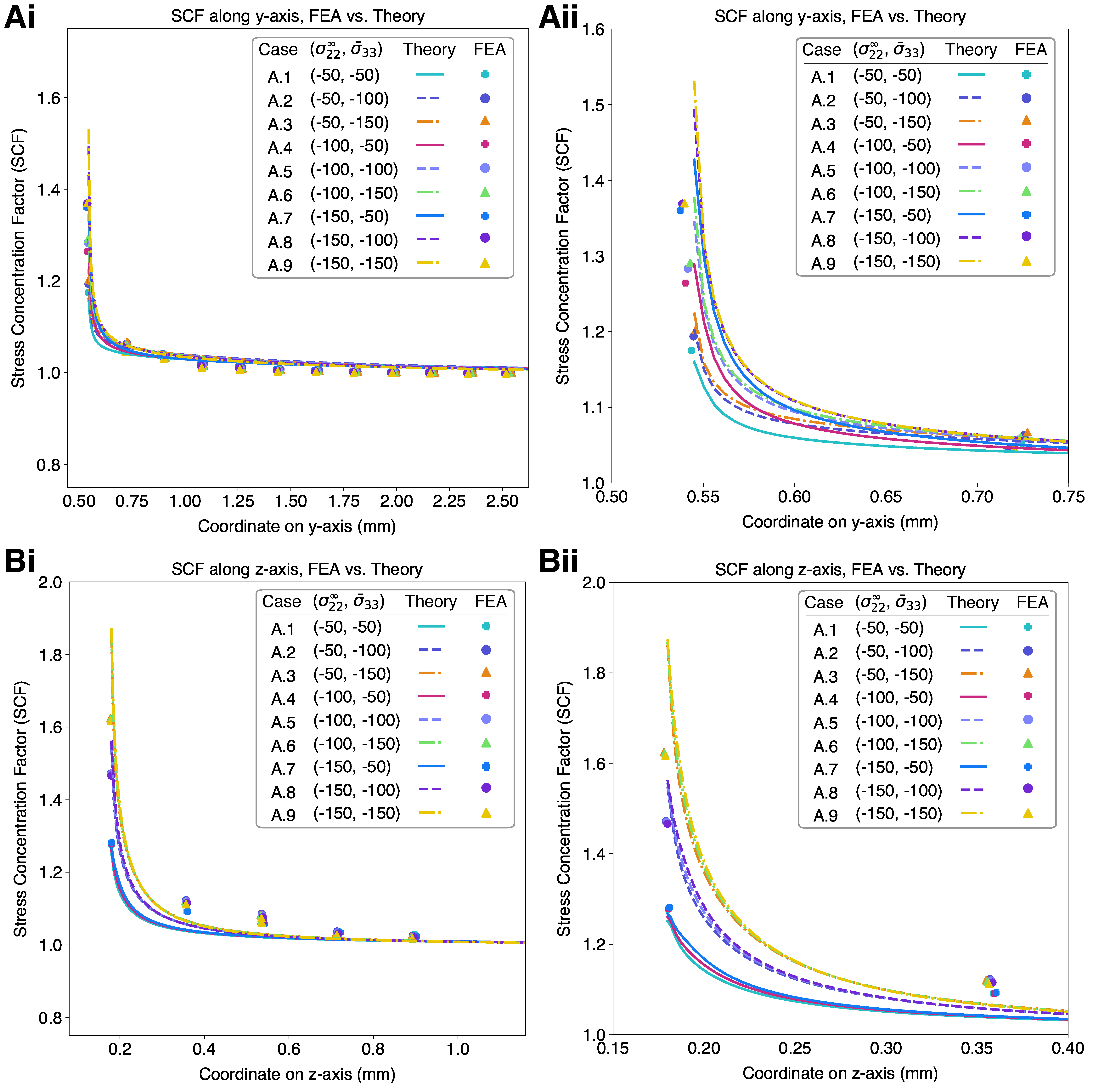}
\caption{Comparison between FEA and theory on the predicted stress concentration magnitude along the positive y-axis and z-axis. Markers are from the element nodal values in FEA, continuous lines are predictions by theory. (Ai) SCF along y-axis covering two adjacent tow widths. (Aii) Magnified view of comparison in (Ai) showing SCF close to the broken tow (about a quarter of tow width). (Bi) SCF along the z-axis (stacking direction). (Bii) Magnified view of data in (Bi) showing SCF close to the broken tow.}
\label{fig:comparison-SCF-yzAxis}
\end{figure}

Fig.\ref{fig:comparison-SCF-yzAxis} shows comparison between SCFs (predicted by FEA and theory) along y-axis and x-axis  for all cases. As expected, maximum stresses occur in regions closest to the broken tow, and decrease when moving away from the tow breakage, and then eventually drop back to far-field levels. It is also observed that the assigned boundary conditions result in exactly the targeted far-field stress levels (e.g., $\sigma^{(k)}_{11,\infty}= 1000 \text{MPa}$ in fiber orientation), which verifies the accuracy of using velocity-based boundary conditions in FEA to obtain the targeted stress levels for each case.

In addition to the SCFs, Table.\ref{tab:comparison-FEA-theory} shows comparison between the predictions of several important values, max SCF, tow break opening displacement, debonding length, for all cases. In which, the max SCF values (and corresponding errors) are evaluated at closest coordinates to the broken tow ($y=w/2$, $z=h/2$, leftmost data in Fig.\ref{fig:comparison-SCF-yzAxis}), and henceforth the error computed at these two points are also where the maximum error occurs. From Fig.\ref{fig:comparison-SCF-yzAxis} and Table.\ref{tab:comparison-FEA-theory}, it is found that excellent agreement is realized in all cases. The errors tend to increase moderately if one or both of the transverse in-situ stresses ($\sigma^{(k)}_{22,\infty}$, $\sigma^{(k)}_{33,\infty}$) are higher. Note that the large transverse in-situ stress values (100MPa, 150MPa for $\sigma^{(k)}_{33,\infty}$) are picked to validate the robustness of theory and FEA. Whereas in practice, typical magnitudes of inner pressures for COPVs are less than 10000 Psi ($\approx$ 72MPa).

\renewcommand{\arraystretch}{1.25}
\begin{table}[!htb]
\centering
\caption{Numerical case studies on uni-directional plies} \label{tab:comparison-FEA-theory}
\begin{threeparttable}
\begin{tabular}{c c cc c c c c c}
    \toprule
    Case & \multicolumn{3}{c}{In-situ stress states (MPa)} & & \multicolumn{2}{c}{Max SCF} & \makecell{Break \\ opening} & \makecell{Ratio between \\ debonding lengths} \\[1ex] \cmidrule(lr){2-4} \cmidrule(lr){6-7} \cmidrule(lr){8-8} \cmidrule(lr){9-9}
    & $\sigma^{(k)}_{11,\infty}$ & $\sigma^{(k)}_{22,\infty}$ & $\bar{\sigma}^{(k)}_{33}$ & & Intra-ply & Inter-ply & 2$\delta_{cod}$ & $L_{\lra}/L_{\uda}$ \\[1ex]
    \hline 
    \multirow{3}{*}{A.1} & \multirow{3}{*}{1000} & \multirow{3}{*}{-50} & \multirow{3}{*}{-50} & Theory & 1.160 & 1.251 & 0.0348 & 0.952 \\
    & & & & FEA & 1.175 & 1.277 & 0.0352 & 0.949 \\ 
    & & & & Difference & 1.30$\%$ & 2.09$\%$ & 1.05$\%$ & -0.28$\%$ \\[1ex]
    \hline 
    \multirow{3}{*}{A.2} & \multirow{3}{*}{1000} & \multirow{3}{*}{-50} & \multirow{3}{*}{-100} & Theory & 1.203 & 1.548 & 0.0191 & 0.913 \\
    & & & & FEA & 1.194 & 1.472 & 0.0201 & 1.00 \\ 
    & & & & Difference & -0.77$\%$ & -4.92$\%$ & 5.00$\%$ & 9.56$\%$ \\[1ex]
    \hline 
    \multirow{3}{*}{A.3} & \multirow{3}{*}{1000} & \multirow{3}{*}{-50} & \multirow{3}{*}{-150} & Theory & 1.225 & 1.861 & 0.0132 & 0.873 \\
    & & & & FEA & 1.203 & 1.624 & 0.0148 & 1.030 \\ 
    & & & & Difference & -1.81$\%$ & -12.70$\%$ & 12.16$\%$ & 17.92$\%$ \\[1ex]
    \hline 
    \multirow{3}{*}{A.4} & \multirow{3}{*}{1000} & \multirow{3}{*}{-100} & \multirow{3}{*}{-50} & Theory & 1.289 & 1.260 & 0.029 & 0.932 \\
    & & & & FEA & 1.264 & 1.278 & 0.0283 & 0.882 \\ 
    & & & & Difference & -1.93$\%$ & 1.40$\%$ & -4.05$\%$ & -5.33$\%$ \\[1ex]
    \hline 
    \multirow{3}{*}{A.5} & \multirow{3}{*}{1000} & \multirow{3}{*}{-100} & \multirow{3}{*}{-100} & Theory & 1.346 & 1.556 & 0.0174 & 0.885 \\
    & & & & FEA & 1.283 & 1.470 & 0.0176 & 0.857 \\ 
    & & & & Difference & -4.66$\%$ & -5.53$\%$ & 1.13$\%$ & -3.19$\%$ \\[1ex]
    \hline 
    \multirow{3}{*}{A.6} & \multirow{3}{*}{1000} & \multirow{3}{*}{-100} & \multirow{3}{*}{-150} & Theory & 1.377 & 1.868 & 0.0123 & 0.838 \\
    & & & & FEA & 1.291 & 1.622 & 0.0133 & 0.896 \\ 
    & & & & Difference & -6.26$\%$ & -13.16$\%$ & 7.78$\%$ & 6.98$\%$ \\[1ex]
    \hline 
    \multirow{3}{*}{A.7} & \multirow{3}{*}{1000} & \multirow{3}{*}{-150} & \multirow{3}{*}{-50} & Theory & 1.377 & 1.868 & 0.0123 & 0.838 \\
    & & & & FEA & 1.291 & 1.622 & 0.0133 & 0.896 \\ 
    & & & & Difference & -4.67$\%$ & 1.09$\%$ & -5.73$\%$ & -9.81$\%$ \\[1ex]
    \hline 
    \multirow{3}{*}{A.8} & \multirow{3}{*}{1000} & \multirow{3}{*}{-150} & \multirow{3}{*}{-100} & Theory & 1.494 & 1.563 & 0.0160 & 0.791 \\
    & & & & FEA & 1.369 & 1.466 & 0.0158 & 0.796 \\ 
    & & & & Difference & -8.33$\%$ & -6.17$\%$ & -0.91$\%$ & 0.66$\%$ \\[1ex]
    \hline 
    \multirow{3}{*}{A.9} & \multirow{3}{*}{1000} & \multirow{3}{*}{-150} & \multirow{3}{*}{-150} & Theory & 1.531 & 1.874 & 0.0116 & 0.712 \\
    & & & & FEA & 1.370 & 1.617 & 0.0119 & 0.810 \\ 
    & & & & Difference & -10.56$\%$ & -13.72$\%$ & 2.75$\%$ & 13.86$\%$ \\[1ex]
    \bottomrule
\end{tabular}
\begin{tablenotes}\footnotesize
\item[*] Total number of plies is 13, tow breakage occurs in the center tow in the 7th ply.
\item[*] Errors of max SCF are evaluated at $(x=0, y=w/2, z=0)$ and $(x=0, y=0, z=h/2)$ for intra- and inter-ply domains, respectively, because these points are where maximum errors occur.
\end{tablenotes}
\end{threeparttable}
\end{table}

\subsection{Summary}

In this section, we performed high-fidelity finite element simulations to evaluate the stress concentration profile around a tow break under different stress states, and compared the numerical results with the predictions by the derived theories. Overall, excellent agreement in essential metrics are observed (Table \ref{tab:comparison-FEA-theory}). It should be emphasized that neither FEA results nor analytical predictions should be considered as the 'correct reference values' because closed-form solutions are arguably not attainable, and henceforth under this circumstance, quantitatively close agreement between these two approaches validate the effectiveness of the derived algorithms.

Despite the transverse in-situ stresses are negative in all cases in this study, there are no specific restrictions on the sign and magnitude of them in the analytical algorithms. For example, if there is zero intra-ply frictions after debonding, then by theory, virtually all of the overload will be distributed into the adjacent plies through inter-ply frictional forces. Such a 'zero slipping force' scenario may not be practical since the filament overwrap should remain as an integrated part until destructive failure. Also, after debonding, the frictional forces may not necessarily be linearly proportional to the contact pressure. It may also depend on other factors such as surface roughness, stacking layup angles, etc. For example, it is observed in an experimental study (also Toray T1100) that the inter-tow angles directly affect the frictional behaviors \cite{Tourlonias.etal2019}. 

In specific, the primary reason for using negative transverse stresses in FEA is to ensure numerical stability. Since the interfaces are represented by the small-thickness cohesive elements embedded between continuum tows, zero or positive in-situ transverse stresses (on the laminate boundaries) will lead to splitting between tows, which would further result in local instabilities. This is because the deformation mechanism of the cohesive elements is based on the relative displacements between pairs of nodes on its top and bottom surfaces, and subsequently, a cohesive element having both positive normal separation and shear deformation would tend to rotate and distort, which leads to dynamic effects and stress oscillations across the debonded interfaces. Non-negative interfacial friction in this study also requires that the top and bottom surfaces of a cohesive element is in contact with negative normal forces, and henceforth loading cases with zero or positive $\sigma^{(k)}_{22,\infty}, \sigma^{(k)}_{33,\infty}$ are not investigated in the finite element simulations. To alleviate the dynamic effect caused by distortion of cohesive elements, specially modified cohesive zone modelling techniques may be required, which is subjective to future studies.

\FloatBarrier

\section{Conclusion}\label{sec:conclusion}
In this study, we derived an empirical analytical approach to predict the overload profile around a broken tow in FWC structures. This kind of problem is challenging due to complicated geometries, interfacial debonding and sliding between domains, and also the load-redistribution mechanisms under different stress states. To resolve this, we firstly applied a shear-lag analysis to solve for tow break opening, perturbation axial displacement of the broken tow, intra- and inter-ply debonding lengths (Section \ref{sec:shearLagAnalysis}). Subsequently, based on the first principals of the load-sharing mechanisms between different domains, we derived formulas that approximate the resultant overload stresses around the broken tow under assumed in-situ stress levels (Section \ref{sec:quantifySCF}). Thereafter, we compared the analytical predictions of critical SCFs under different stress states to the values predicted by high-fidelity finite element analysis, and observed excellent agreement between them (Section \ref{sec:feaValidation}).

The derived algorithms conveniently predict the overload stress magnitude at a location of interest in FWCs, and henceforth, could be used as a complementary evaluation at meso scale to improve theoretical statistical strengths prediction of such composites. On the other side, the algorithms could be implemented in development of failure criteria for continuum elements in component-scale finite element models. Also, this framework may be adapted to structures with similar formations, such as tape-based composites, layered structures printed by direct ink writing, etc.

There are also limitations in this study. Firstly, the representative cross-section of tows is assumed to be rectangular, whereas in practice it can be somewhere between oval to rectangular. Secondly, the characterization of resin-rich areas between tows, despite being straightforwardly observable from failure pattern of FWC structures, is somehow empirical, and may need to be further verified by experimental and tomographic studies. Thirdly, due to the potential numerical instabilities in debonded interfaces (cohesive elements) when the normal strain is zero or positive, the in-situ transverse stresses in the case studies are set to negative values (Table.\ref{tab:comparison-FEA-theory}). That said, the effectiveness of the derived algorithms have been verified by varying of the stresses states, and thus the algorithms can be used for general stress states despite lack of corresponding numerical validation at the moment. Moreover, the interfacial response is characterized as linear traction-separation before debonding, followed by pressure-dependent sliding after debonding (Fig.\ref{fig:debond-and-slip}, while in practice epoxy resin interfaces can exhibit certain extent of non-linearity (before debonding). Lastly, inter-ply overload magnitudes for FWCs with general stacking layup angles are predicted by applying multiplication of trigonometric terms to the results on uni-directional tows (Eq.\ref{eq:SCF_zAxis_local}, which may need to be verified by additional numerical studies in the future. 

Potential subsequent work can be categorized into two directions. The first is specific applications of the proposed analytical algorithms, including the aforementioned development of a theoretical framework for statistical strength prediction of filament-wound composites. The second direction involves refinement and additional validation of the present study, such as stress concentration profiles between plies with different winding angles.

\section*{Acknowledgments}
A new empirical analytical approach is developed for predicting the stress concentration profile around an in-situ tow break in filament-wound composites. A shear-lag analysis is firstly performed to solve for the perturbational axial displacement of the broken tow and resultant debonding lengths. Solution of stress field caused by tangential load on the surface of a elastic half space is then utilized in combination with superposition concepts to obtain the overload magnitudes in the neighboring tows. Subsequently, high-fidelity finite element analysis on a representative uni-directional laminate model under different stress states is performed, and excellent overall agreement is observed between analytical and numerical predictions. The proposed method takes into account essential aspects such as transversely isotropic material properties, in-situ stress states and their effect on the interfacial frictional forces in the debonded interfaces, and thus provides a convenient way to evaluate the stress concentration factors in damaged filament-wound composites. In addition, this approach can be applied to yield auxiliary failure evaluation criteria for statistical strength prediction or finite element modeling of filament-wound composites or similar structures. The authors acknowledge funding support from research grant provided by National Institute of Standards and Technology (NIST) under agreement ID: 70NANB14H323. The first author also had received support from Cornell Graduate School.

\section*{Data availability}
Code for analytical predictions, input files (.inp) for FEA (Abaqus/Explicit), and corresponding user subroutines (VUMAT) used for this study are available from the authors upon request.

{
\footnotesize
\bibliography{SCF_M1_Reference}
}

\listoffigures
\listoftables

\appendix

\section{Derivation of solutions in shear-lag analysis}\label{sec:appendix-solution-procedure}
The steps to obtain solutions are straightforward, and are the same for all three cases, with Case 3 ($L_{\lra} = L_{\uda} $) requiring less algebras. Therefore, only steps for Case 1 are briefly explained here.

As shown in Eq.\ref{eq:condition_for_3_cases}, Case 1 means that:
\begin{equation}\label{eq:append_condtion_Case1}
L_{\lra} > L_{\uda}  \:, \qquad \frac{\tau^f_{s,\lra}}{K_{s,\lra}} < \frac{\tau^f_{s,\uda}}{K_{s,\uda}} 
\end{equation}
And from the governing equations, the general solution takes form:
\begin{equation}\label{eq:append_Case1_general_sol}
    \delta u^{(k,0)}_{11}(x) = \left\{\begin{array}{ll}
    \dfrac{A_2 + A_3}{2 A_1} \cdot x^2 + C_1 \cdot x + C_2 \:, &  x \in \left[0\:,L_{\uda} \right) \\
    -A_2/A_5 + C_3 \cdot \exp \left(\sqrt{A_5/A_1}\cdot x \right) + C_4 \cdot \exp \left(-\sqrt{A_5/A_1}\cdot x \right)   \:, &  x \in \left[L_{\uda} \:, L_{\lra} \right) \vpb\\
    C_5 \cdot \exp \left(\sqrt{(A_4 + A_5)/A_1 }\cdot x \right) + C_6 \cdot \exp \left(-\sqrt{(A_4 + A_5)/A_1 }\cdot x \right) \:, &  x \in \left[L_{\lra} \:, +\infty \right) \vpb
\end{array}\right.
\end{equation}
where $A_1$ to $A_5$ are non-negative constants involving material properties and stress states (as defined in Eq.\ref{eq:A1_to_A5_definition}). The eight boundary conditions (BCs) are listed in Eq.\ref{eq:Case1_BCs} with sub-equations numbered from 1 to 8. Here denote them as BC-1 to BC-8 for simplicity. Then BC-1 and BC-2 directly result in $C_1 = -\sigma^{(k)}_{11,\infty}/E_l$ and $C_5 =0$, whereby solution is simplified to the form shown in Eq.\ref{eq:Case_1_general_solution}. 

Substitute BC-8 into BC-4 and BC-6, respectively, then two of the remaining undetermined constants can be expressed in terms of $L_{\lra}$ as:
\begin{equation}\label{eq:append_C3_C4}
C_3 = \frac{A_7}{2 A_5 K_{s,\lra}} \cdot \exp\left(-\sqrt{\frac{A_5}{A_1}} \cdot L_{\lra} \right) \,, \quad
C_4 = \frac{A_6}{2 A_5 K_{s,\lra}} \cdot \exp\left(\sqrt{\frac{A_5}{A_1}} \cdot L_{\lra} \right)	
\end{equation}  
where 
\begin{equation}\label{eq:append_A6_A7}
A_6 = A_2 K_{s,\lra} + \left(A_5 + \sqrt{A_4 A_5 + A_5^2}\right)\tau^f_{s,\lra} \,, \quad
A_7 = A_2 K_{s,\lra} + \left(A_5 - \sqrt{A_4 A_5 + A_5^2}\right)\tau^f_{s,\lra}
\end{equation}
Then substitute Eq.\ref{eq:append_C3_C4} into BC-3 and BC-7, the resulting two equations, BC-3* and BC-7* can be expressed in terms of $L_{\uda}$ and $(L_{\uda}-L_{\lra})$ as 
\begin{equation}\label{eq:appedn_BC3_7_new}
	\begin{aligned}
	&\text{BC-3*:} \quad  \frac{A_2 + A_3}{A_1}E_l \cdot L_{\uda} +   \frac{E_l \sqrt{A_5/A_1}}{2 A_5 K_{s,\lra}}\cdot \left[  A_6 \exp \left(-\sqrt{\frac{A_5}{A_1}} \cdot L_{\Delta} \right) - A_7 \exp \left(\sqrt{\frac{A_5}{A_1}} \cdot L_{\Delta} \right)  \right] - \sigma^{(k)}_{11,\infty}  = 0  \\
	&\text{BC-7*:} \quad \frac{K_{s,\uda}}{2A_5 K_{s,\lra}}\left[ A_6 \exp \left(-\sqrt{\frac{A_5}{A_1}} \cdot L_{\Delta} \right) + A_7 \exp \left(\sqrt{\frac{A_5}{A_1}} \cdot L_{\Delta} \right) \right] -\frac{A_2}{A_5} K_{s,\uda} -\tau^f_{s,\uda} = 0 
	\end{aligned}
\end{equation}
where
\begin{equation}\label{eq:append_Delta_L}
	L_{\Delta} \equiv L_{\uda} - L_{\lra} <0 
\end{equation}
as shown before, is the difference between two debonding lengths (negative in Case 1).

It is observed that components involving $L_{\Delta}$ are exponential terms whose powers have the same magnitude and opposite signs ($\sqrt{A_5/A_1}L_{\Delta}$), therefore, the following operations can be applied to cancel out the terms containing $L_{\Delta}$:
\begin{equation}\label{eq:append_two_operations_to_cancel_out_DeltaL}
\begin{aligned}
	& \left[ (\text{BC-7*} )\cdot \sqrt{\frac{A_5}{A_1}} \cdot \frac{E_l}{K_{s,\uda}} +  (\text{BC-3*}) \right] = 0  \quad \longrightarrow \quad \exp \left(-\sqrt{\frac{A_5}{A_1}}\cdot L_{\Delta} \right) = \Psi(L_{\uda}) >0 \vpb \\
	& \left[ (\text{BC-7*} )\cdot \sqrt{\frac{A_5}{A_1}} \cdot \frac{E_l}{K_{s,\uda}} -  (\text{BC-3*})  \right] = 0 \quad  \longrightarrow \quad \exp \left(\sqrt{\frac{A_5}{A_1}}\cdot L_{\Delta} \right) = \Upsilon(L_{\uda}) >0 \vpb 
\end{aligned}
\end{equation}
where $\Psi(L_{\uda})$ and $\Upsilon(L_{\uda})$ are functions of variable $L_{\uda}$ and other constants, the detailed expressions are:
\begin{equation}\label{eq:expression_Psi_Upsilon}
\begin{aligned}
    \Psi(L_{\uda}) &= -\dfrac{A_2 + A_3}{A_6}\sqrt{\dfrac{A_5}{A_1}} K_{s,\lra} \cdot L_{\uda} + \dfrac{K_{s,\lra}}{A_6}\left( A_2 + A_1 \sqrt{\frac{A_5}{A_1}} \frac{\sigma^{(k)}_{11,\infty}}{E_l} + A_5 \frac{\tau^f_{s,\uda}}{K_{s,\uda}}\right)  \\
    \Upsilon(L_{\uda}) &= \dfrac{A_2 + A_3}{A_7}\sqrt{\dfrac{A_5}{A_1}} K_{s,\lra} \cdot L_{\uda} + \dfrac{K_{s,\lra}}{A_7}\left( A_2 - A_1 \sqrt{\frac{A_5}{A_1}} \frac{\sigma^{(k)}_{11,\infty}}{E_l} + A_5 \frac{\tau^f_{s,\uda}}{K_{s,\uda}}\right)
\end{aligned}
\end{equation}
Then the above two relations can be simplified into a quadratic equation of $L_{\uda}$ through:
\begin{equation}\label{eq:append_quadratic_eq_for_L_uda}
\Psi(L_{\uda}) \cdot \Upsilon(L_{\uda}) = 1 \quad \longrightarrow \quad A_8 \cdot L_{\uda}^2  +  A_9 \cdot L_{\uda} + A_{10} = 0
\end{equation}
where 
\begin{equation}\label{eq:appedn_def_A8910}
\begin{aligned}
    & A_8 = \left( \frac{E_l}{A_1} \right)^2 \cdot (A_2 + A_3)^2  \geq 0\, \qquad A_9 = -\frac{2 E_l}{A_1} \sigma^{(k)}_{11,\infty} \cdot (A_2 + A_3) \leq 0 \vpb \\
    & A_{10} = \left(\sigma^{(k)}_{11,\infty} \right)^2 - 
    \frac{A_2 E_l^2}{A_1} \left( \frac{A_2}{A_5} + 2\frac{\tau^f_{s,\uda} }{K_{s,\uda}} \right) 
    + \frac{E_l^2}{A_1} \left[ \frac{A_6 A_7}{A_5 K_{s,\lra}^2} - A_5\left( \frac{\tau^f_{s,\uda} }{K_{s,\uda}} \right)^2 \right]
\end{aligned}	
\end{equation}
are constants, and therefore the two solutions for $L_{\uda}$ are:
\begin{equation}\label{eq:append_sol_Luda_1and2}
	L_{\uda}^a = \dfrac{-A_9 - \sqrt{A_9^2 - 4 A_8 A_{10}} }{2 A_8} \,, \qquad L_{\uda}^b = \dfrac{-A_9 + \sqrt{A_9^2 - 4 A_8 A_{10}} }{2 A_8}
\end{equation}
Mathematically, in order for the solution pair to be real and positive, the required conditions are:
\begin{equation}\label{eq:append_requirement_for_real_positive_sol}
	(A_2 + A_3) > 0 \,, \qquad A_{10} > 0
\end{equation}
which directly lead to the formulas given in Eq.\ref{eq:A2+A3>0} and Eq.\ref{eq:min_vlaue_for_siginfy}, respectively. In a physical perspective, as explained in Section \ref{subsec:shearlag-solution}, these two conditions ensure that: 1) there are interface debonding happening, so that the shear-lag model is valid; 2) there are non-zero frictional shear forces in the debonded interfaces, so that the broken tow will not result in unstable debonding, i.e., the system will reach an equilibrium state.

In the next appendix section, it will be shown that the smaller one of the pair, $L_{\uda}^a$, is the only valid solution, since the other one is against the prerequisite definition for each case. Then, the remaining unused boundary condition, BC-5, lead to solution of the last unknown constant, $C_2$. In brief, by solving for the two debonding lengths ($L_{\uda}, L_{\lra}$), all the other undetermined constants can be solved accordingly.

\section{Proof of the uniqueness of solutions}\label{sec:appendix-proof-of-uniqueness}

Following the derivation of solutions for Case 1 as shown in \ref{sec:appendix-solution-procedure}, we continue to show that the solution is uniquely determined. Proofs for the other two cases are of the same steps and thus they are skipped here. 

From the constraints in Eq.\ref{eq:append_two_operations_to_cancel_out_DeltaL}, the relation between $L_{\uda}$ and $L_{\Delta}$ can be obtained by:
\begin{equation}\label{eq:append_relatino_between_L_uda_Lc}
	\dfrac{\Upsilon(L_{\uda})}{\Psi(L_{\uda})} = \exp \left(2 \sqrt{\frac{A_5}{A_1}} \cdot L_{\Delta} \right) = \dfrac{A_6}{A_7} \cdot \dfrac{A_1 B_3 + B_1 L_{\uda}}{A_1 B_2 - B_1 L_{\uda}} \equiv \Phi(L_{\uda})  
\end{equation}
where the three new constants are:
\begin{equation}\label{eq:appedn_def_B123}
\begin{aligned}
& B_1 = (A_2 + A_3) A_5 E_l K_{s,\uda} \\
& B_2 = A_2 E_l K_{s,\uda} \sqrt{\frac{A_5}{A_1}} + A_5 K_{s,\uda}\sigma^{(k)}_{11,\infty} + A_5 \sqrt{\frac{A_5}{A_1}} \tau^f_{s,\uda} E_l  \,, \\
& B_3 = A_2 E_l K_{s,\uda} \sqrt{\frac{A_5}{A_1}} - A_5 K_{s,\uda} \sigma^{(k)}_{11,\infty} + A_5 \sqrt{\frac{A_5}{A_1}} \tau^f_{s,\uda} E_l
\end{aligned}	
\end{equation}

Then the two candidate solutions of $L_{\uda}$, denoted as $L_{\uda}^a, L_{\uda}^b$ in Eq.\ref{eq:append_sol_Luda_1and2}, will lead to two corresponding values of $L_{\Delta}$ as:
\begin{equation}\label{eq:relation_Luda_DeltaL}
\exp \left(2 \sqrt{\frac{A_5}{A_1}} \cdot L_{\Delta}^a \right) = \Phi(L_{\uda}^a) \,, \quad \exp \left(2 \sqrt{\frac{A_5}{A_1}} \cdot L_{\Delta}^b \right) = \Phi(L_{\uda}^b)
\end{equation}
based on their relation as shown in Eq.\ref{eq:append_relatino_between_L_uda_Lc}.

Now we prove that the smaller one of the two solutions, $L^a_{\uda}$, is the only valid one, because the other solution, $L^b_{\uda}$, will lead to a positive $L_{\Delta}$ which is against the definition (Eq.\ref{eq:Case1_debond_lengths} or \ref{eq:append_Delta_L}) for this case. Firstly, it is verified that:
\begin{equation}\label{eq:append_prove_DeltaL2>DeltaL1}
\dfrac{\exp \left(2 \sqrt{\frac{A_5}{A_1}} \cdot L_{\Delta}^b \right) }{\exp \left(2 \sqrt{\frac{A_5}{A_1}} \cdot L_{\Delta}^a \right)} -1 = \dfrac{\Phi(L^b_{\uda})}{\Phi(L^a_{\uda}) } - 1 = \dfrac{A_1 B_1 \left(B_2 + B_3 \right) \left(L_{\uda}^b-L_{\uda}^a \right) }{\left(A_1 B_3 + B_1 L_{\uda}^a \right) \left(A_1 B_2 - B_1 L_{\uda}^b \right) } >0
\end{equation}
so that $L_{\Delta}^b > L_{\Delta}^a$, because $e^{2\sqrt{A_5/A_1}\cdot t }$ is a positive and strictly increasing function for any $t$.
To verify Eq.\ref{eq:append_prove_DeltaL2>DeltaL1}, it is readily seen that
\begin{equation}\label{eq:append_numerator}
	A_1 > 0 \,, \quad B_1 > 0 \,, \quad B_2 + B_3 > 0 \,, \quad L_{\uda}^b-L_{\uda}^a = -L_{\Delta} > 0
\end{equation}
so the numerator is positive. Meanwhile, it can be verified that the denominator is also positive, namely,
\begin{equation}\label{eq:append_denomitor}
\begin{aligned}
& \left(A_1 B_3 + B_1 L_{\uda}^a \right) \left(A_1 B_2 - B_1 L_{\uda}^b \right) \\ 
& \quad = A_1 A_5 E_l^2 \left[\sqrt{ \left( \dfrac{A_2 K_{s,\uda} + A_5 \tau^f_{s,\uda} }{K_{s,\lra}} \right)^2 - A_6 A_7 \left(\dfrac{K_{s,\uda}}{K_{s,\lra}}\right) ^2 }  -\left(A_2 K_{s,\uda} +A_5 \tau^f_{s,\uda} \right)  \right]^2 >0
\end{aligned}
\end{equation}
because both $A_1$ and $A_5$ are positive, and the remaining are squared terms.
Secondly, we show that 
\begin{equation}\label{eq:append_prove_Delta2Delta1_pos_neg}
\left[\exp \left(2 \sqrt{\frac{A_5}{A_1}} \cdot L_{\Delta}^a \right) -1 \right]\cdot \left[ \exp \left(2 \sqrt{\frac{A_5}{A_1}} \cdot L_{\Delta}^b \right) -1 \right] = \left[\Phi(L^a_{\uda}) -1 \right]\cdot \left[ \Phi(L^b_{\uda}) -1 \right] < 0
\end{equation}
so that $L_{\Delta}^a$ and $L_{\Delta}^b$ have opposite signs, because $\exp(0) = 1$ and $e^{2\sqrt{A_5/A_1}\cdot t}$ is a positive, strictly increasing function for any $t$. This way, $L_{\Delta}^b$ must be positive since it is already proved that $ L_{\Delta}^b > L_{\Delta}^a $. However, $L_{\Delta}$ must be a negative value by its definition for this case, therefore, the solution of all unknowns are uniquely determined by $L_{\uda}^a$ and the other candidate solution of inter-ply debonding length, $L_{\uda}^b$, is invalid.
To verify Eq.\ref{eq:append_prove_Delta2Delta1_pos_neg}, it can be shown that
\begin{equation}\label{eq:append_prove_Delta2Delta1_pos_neg_2}
\begin{aligned}
	& \left[\Phi(L^a_{\uda}) -1 \right]\cdot \left[ \Phi(L^b_{\uda}) -1 \right] \\
	& \quad = \left[ \dfrac{4A_5 K_{s,\lra} \left( 2 A_2 K_{s,\uda}K_{s,\lra} + A_5 K_{s,\lra} \tau^f_{s,\uda} + A_5 K_{s,\uda} \tau^f_{s,\lra} \right) }{ K_{s,\uda} \left( A_2 K_{s,\lra} + (A_5 - \sqrt{A_4 A_5 + A_5^2})\tau^f_{s,\lra} \right)^2 }  \right] \cdot \left[ \frac{\tau^f_{s,\lra}}{K_{s,\lra}} - \frac{\tau^f_{s,\uda}}{K_{s,\uda}} \right]
\end{aligned}
\end{equation}
where it is a multiplication of two terms, it is obvious that the first term is positive, and by the definition of Case 1 as shown in Eq.\ref{eq:condition_for_3_cases}
\begin{equation}
\text{Case 1:} \quad L_{\lra} > L_{\uda} \quad \longrightarrow \quad \tau^f_{s,\lra}/K_{s,\lra} < \tau^f_{s,\uda}/ K_{s,\uda} \quad \longrightarrow \quad \frac{\tau^f_{s,\lra}}{K_{s,\lra}} - \frac{\tau^f_{s,\uda}}{K_{s,\uda}} < 0
\end{equation}
so that Eq.\ref{eq:append_prove_Delta2Delta1_pos_neg} holds true.

Therefore, it is proved that the smaller one of the two solutions for the inter-debonding length, $L_{\uda}^a$ in Eq.\ref{eq:append_sol_Luda_1and2} is the only valid solution. From which, $L_{\Delta}$ is uniquely given by Eq.\ref{eq:append_relatino_between_L_uda_Lc} so that both debonding lengths can be determined (Eq.\ref{eq:Case1_debond_lengths}). Then the remaining undetermined constants can be expressed in terms of the two debonding lengths as given in Eq.\ref{eq:Case1_constants_Ci}.

\section{Solution for Case 2}\label{sec:appendix-sol-for-case2}
As shown in Eq.\ref{eq:condition_for_3_cases}, the condition for $L_{\lra} < L_{\uda}$ is:
\begin{equation}
\text{Case 2:} \qquad \tau^f_{s,\lra}/K_{s,\lra} > \tau^f_{s,\uda}/ K_{s,\uda}
\end{equation}
Similar to Case 1 in the main text, by setting $\delta u^{(k,0)}_{11} (x) \equiv \delta u$ for simplicity, and by utilizing the previously defined constants (Eq.\ref{eq:A1_to_A5_definition}), the governing equations for Case 2 can be expressed in form:
\begin{subnumcases}{\dfrac{d^2}{dx^2} \delta u =}
\frac{A_2 + A_3}{A_1},  & $ x \in \rinterval{0}{L_{\lra}} \vpb $ \label{eq:GE.Case2.interv1} \\
\frac{1}{A_1} \left( A_3 + A_4 \cdot \delta u \right),  & $ x \in \rinterval{L_{\lra}}{L_{\uda}} \vpb $ \label{eq:GE.Case2.interv2} \\
\frac{A_4 + A_5}{A_1} \delta u,  & $ x \in \rinterval{L_{\uda}}{+\infty} \vpb $ \label{eq:GE.Case2.interv3}
\end{subnumcases}
and eight boundary conditions are comparable to those shown in Eq.\ref{eq:Case1_BCs}, but the expression differ due to change of intervals.

After applying the similar procedures, the debonding lengths (for Case 2) can be solved:
\begin{equation}\label{eq:Case2_debond_lengths}
\begin{aligned}
L_{\lra} & = -\dfrac{1}{2A_8} \left( A_9 + \sqrt{A_9^2- 4 A_8 A^{[2]}_{10}}  \right) >0 \,,\vpb \\
L_{\Delta}  & \equiv L_{\uda} - L_{\lra} =  -\frac{1}{2} \sqrt{A_1/A_4} \cdot \ln \left(\dfrac{A_6^{[2]}}{A_7^{[2]}} \cdot \dfrac{A_1 B_3^{[2]} + B_1^{[2]} L_{\lra}}{A_1 B_2^{[2]} - B_1^{[2]} L_{\lra}} \right) > 0 \,,\vpb \\
L_{\lra} & =  L_{\uda} - L_{\Delta} < L_{\uda} \vpb
\end{aligned}
\end{equation}
where the constants without the superscript $[2]$ indicates that the expressions remains identical (case independent), and the updated constants specific to Case 2 are:
\begin{equation}\label{eq:append_A6_A7_A10_for_case2}
\begin{aligned}
A_6^{[2]} &= A_3 K_{s,\uda} + \left(A_4 + \sqrt{A_4 A_5 + A_4^2}\right)\tau^f_{s,\uda} \\
A_7^{[2]} &= A_3 K_{s,\uda} + \left(A_4 - \sqrt{A_4 A_5 + A_4^2}\right)\tau^f_{s,\uda} \\
A_{10}^{[2]} &= \left(\sigma^{(k)}_{11,\infty} \right)^2 - 
    \frac{A_3 E_l^2}{A_1} \left( \frac{A_3}{A_4} + 2\frac{\tau^f_{s,\lra} }{K_{s,\lra}} \right) 
    + \frac{E_l^2}{A_1} \left[ \frac{A_6^{[2]} A_7^{[2]}}{A_4 K_{s,\uda}^2} - A_4\left( \frac{\tau^f_{s,\lra} }{K_{s,\lra}} \right)^2 \right]
\end{aligned}
\end{equation}
and
\begin{equation}\label{eq:appedn_def_B123_case2}
\begin{aligned}
& B_1^{[2]} = (A_2 + A_3) A_4 E_l K_{s,\lra} \\
& B_2^{[2]} = A_3 E_l K_{s,\lra} \sqrt{\frac{A_4}{A_1}} + A_4 K_{s,\lra}\sigma^{(k)}_{11,\infty} + A_4 \sqrt{\frac{A_4}{A_1}} \tau^f_{s,\lra} E_l  \,, \\
& B_3^{[2]} = A_3 E_l K_{s,\lra} \sqrt{\frac{A_4}{A_1}} - A_4 K_{s,\lra} \sigma^{(k)}_{11,\infty} + A_4 \sqrt{\frac{A_4}{A_1}} \tau^f_{s,\lra} E_l
\end{aligned}
\end{equation}

\section{Solution for Case 3}\label{sec:appendix-sol-for-case3}
In Case 3, the debonding lengths are equivalent, henceforth the intervals for the piecewise governing equations of the perturbation axial displacement of the broken tow reduces to two, and the number of boundary constraints and undetermined parameters reduce to six.

Set $L_{\lra} = L_{\uda} = L_d$, the governing equations for Case 3 are:
\begin{subnumcases}{\dfrac{d^2}{dx^2} \delta u =}
\frac{A_2 + A_3}{A_1},  & $ x \in \rinterval{0}{L_d} \vpb $ \label{eq:GE.Case3.interv1} \\
\frac{A_4 + A_5}{A_1} \delta u,  & $ x \in \rinterval{L_d}{+\infty} \vpb $ \label{eq:GE.Case3.interv2}
\end{subnumcases}
The solutions are
\begin{subnumcases}{\delta u =}
\frac{A_2 + A_3}{2A_1}x^2 - \frac{\sigma^{(k)}_{11,\infty}}{E_l}x + C_2,  & $ x \in \rinterval{0}{L_d} \vpb $ \label{eq:SOL.Case3.interv1} \\
\frac{A_4 + A_5}{A_1} \delta u,  & $ x \in \rinterval{L_d}{+\infty} \vpb $ \label{eq:SOL.Case3.interv2}
\end{subnumcases}
where the debonding lengths are:
\begin{equation}\label{eq:case3_debond_L}
    L_d = L_{\lra} = L_{\uda} = \frac{A_1}{A_2 + A_3}\left( \frac{\sigma^{(k)}_{11,\infty}}{E_l} - \sqrt{\frac{A_4 + A_5}{A_1}}\frac{\tau^f_{s,\lra}}{K_{s,\lra}} \right)
\end{equation}
and the expression of the unknown constant $C_2$ is:
\begin{equation}\label{eq:c2_in_case3}
\begin{aligned}
    C_2 &= -\frac{A_2 + A_3}{2A_1}L_d^2 + \frac{\sigma^{(k)}_{11,\infty}}{E_l} L_d + \frac{\tau^f_{s,\lra}}{K_{s,\lra}} \\
    & = \frac{A_1}{2(A_2+ A_3)}\left(\frac{\sigma^{(k)}_{11,\infty}}{E_l}\right)^2 - \frac{A_4 + A_5}{2(A_2 + A_3)}\left(\frac{\tau^f_{s,\lra}}{K_{s,\lra}}\right)^2 + \frac{\tau^f_{s,\lra}}{K_{s,\lra}}
\end{aligned}
\end{equation}

\end{document}